\documentclass[12pt,preprint]{aastex}
\usepackage{emulateapj5}
\font\sevenrm=cmr7 
\def\kmm{{\sevenrm KMM}~}
\shorttitle{A New Population of Radio Quasars}
\shortauthors{Padovani et al.}

\begin{document}

\title{What Types of Jets does Nature Make: A New Population of 
Radio Quasars}

\author{Paolo Padovani\altaffilmark{1}, Eric
S. Perlman\altaffilmark{2}, Hermine Landt\altaffilmark{3}}

\affil{Space Telescope Science Institute, 3700 San Martin Drive, 
Baltimore, MD  21218, USA}

\and

\author{Paolo Giommi, Matteo Perri\altaffilmark{4}}

\affil{ASI Science Data Center, ASDC, 
c/o ESRIN, Via G. Galilei, I-00044 Frascati, Italy}

\altaffiltext{1}{ESA Space Telescope Division}

\altaffiltext{2}{Joint Center for Astrophysics, University of Maryland, 1000 
Hilltop Circle, Baltimore, MD 21250, USA (current address)}

\altaffiltext{3}{Hamburger Sternwarte, Gojenbergsweg 112, D-21029
Hamburg, Germany}

\altaffiltext{4}{Dipartimento di Fisica, Universit\`a la Sapienza, P.le
A. Moro 2, Roma, Italy}

\begin{abstract}
We use statistical results from a large sample of about 500 blazars, based on
two surveys, the Deep X-ray Radio Blazar Survey (DXRBS), nearly complete, and
the RASS-Green Bank survey (RGB), to provide new constraints on the spectral
energy distribution of blazars, particularly flat-spectrum radio quasars
(FSRQ). This reassessment is prompted by the discovery of a population of FSRQ
with spectral energy distribution similar to that of high-energy peaked BL
Lacs. The fraction of these sources is sample dependent, being $\sim 10\%$ in
DXRBS and $\sim 30\%$ in RGB (and reaching $\sim 80\%$ for the {\it Einstein}
Medium Sensitivity Survey). We show that these ``X-ray strong'' radio quasars,
which had gone undetected or unnoticed in previous surveys, indeed are the
strong-lined counterparts of high-energy peaked BL Lacs and have synchrotron
peak frequencies, $\nu_{\rm peak}$, much higher than ``classical'' FSRQ,
typically in the UV band for DXRBS. Some of these objects may be 100 GeV --
TeV emitters, as are several known BL Lacs with similar broadband spectra. Our
large, deep, and homogeneous DXRBS sample does not show anti-correlations
between $\nu_{\rm peak}$ and radio, broad line region, or jet power, as
expected in the so-called ``blazar sequence'' scenario. However, the fact that
FSRQ do not reach X-ray-to-radio flux ratios and $\nu_{\rm peak}$ values as
extreme as BL Lacs and the elusiveness of high $\nu_{\rm peak}$--high-power
blazars suggest that there might be an intrinsic, physical limit to the
synchrotron peak frequency that can be reached by strong-lined, powerful
blazars. Our findings have important implications for the study of jet
formation and physics and its relationship to other properties of active
galactic nuclei.
\end{abstract}


\section {A Phenomenological Description of the Blazar Class}\label{descript}

Blazars are one of the most extreme classes of active galactic nuclei (AGN),
distinguished by high luminosity, rapid variability, high polarization, radio
core-dominance, and apparent superluminal speeds \citep{urr95}. Their
broad-band emission extends from the radio up to the gamma-ray band, and is
dominated by non-thermal processes. The blazar class includes BL Lacertae
objects, characterized by an almost complete lack of emission lines, and the
flat-spectrum radio quasars (FSRQ), which by definition display broad, strong
emission lines.

More than $95\%$ of all known blazars have been discovered either in radio or
X-ray surveys \citep{pad95a}. 
Previous work has shown that X-ray and radio selection methods
yield objects with somewhat different properties at least for BL Lacs. The
energy output of most radio selected BL Lacs peaks in the IR/optical
\citep{gio94,pad95b,pad96}. These objects are more highly polarized
\citep{jan94}, and are more core-dominated in the radio
\citep{per93,lau93,kol96, rec99} and are now referred to as LBL (low-energy
peaked BL Lacs). By contrast, the energy output of most X-ray selected BL Lacs
peaks in the UV/X-ray; these objects, which are referred to as HBL
(high-energy peaked BL Lacs), are less polarized and generally have preferred
position angles of polarization in the optical \citep{jan94}, and are less
core-dominant in the radio \citep{per93,kol96,rec00}.

\citet{pad95b} have demonstrated that the difference
in broad-band peaks for HBL and LBL is not simply
phenomenological. Rather, it represents a fundamental difference
between the two sub-classes. In this respect, one could expect to find
a similar division in the FSRQ class -- for which, until recently, no
evidence existed. Indeed, it was suggested by some authors
\citep{sam96}, based upon the similarities of the optical--X-ray
broad-band spectral characteristics of LBL and FSRQ, that no FSRQ with
synchrotron peak emission in the UV/X-ray band should exist. 

The dichotomy that there existed both high and low energy peaked BL
Lacs (but virtually no intermediate objects), and only low energy
peaked FSRQ was the state of the art in our knowledge of blazar
spectral energy distributions (SEDs) in the mid 1990s. Four
discoveries have drastically changed this picture:

\begin{enumerate}

\item The multi-frequency catalog of \citet{pad97}, which included all 
FSRQ known
before DXRBS and other recent surveys, identified more than 50 FSRQ
($\sim 17\%$ of the FSRQ in their catalog) with broad-band spectra
similar to those of HBL.

\item About 30\% of FSRQ found in the Deep X-ray Radio Blazar Survey (DXRBS;
Perlman et al. 1998; Landt et al. 2001) were found to have X-ray-to-radio
luminosity ratios, $L_{\rm x}/L_{\rm r}$, typical of HBL ($L_{\rm x}/L_{\rm r}
\ga 10^{-6}$ or $\alpha_{\rm rx} \la 0.78$), but broad (full width half maximum 
$> 2,000
{\rm ~km ~s^{-1}}$) and luminous ($L > 10^{43} {\rm ~erg ~s^{-1}}$) emission
lines typical of FSRQ.

\item The discovery of large numbers of ``intermediate'' BL Lacs, objects with
broadband properties intermediate between HBL and LBL, in the DXRBS
\citep{per98,lan01}, ROSAT-Green Bank (RGB; Laurent-Muehleisen et
al. 1998) and other samples \citep{nass96,kock96}.

\item {\it BeppoSAX} observations have shown that the X-ray emission of at
least one FSRQ is dominated by synchrotron radiation, with a peak
frequency $\nu_{\rm peak} \sim 2 \times 10^{16}$ Hz and steep
($\alpha_{\rm x} \sim 1.5$) X-ray spectrum
\citep{pad02}. Two more FSRQ show evidence of $\nu_{\rm peak} \approx 10^{15}$
Hz.

\end{enumerate}

The discovery of ``X-ray strong'' FSRQ (labeled HFSRQ by Perlman et al. 1998
to parallel the HBL moniker; the ``standard'' FSRQ would then be the LFSRQ)
represents a fundamental change in our perception of the broadband spectrum of
FSRQ, similar to that brought about by the X-ray selected BL Lacs discovered
in the {\it Einstein} Medium Sensitivity Survey (EMSS) \citep{mor91,sto91,
rec00}. A preliminary investigation of the properties of HFSRQ in the DXRBS
survey was presented in \citet{per98}. The discovery of a large number of
intermediate BL Lacs should have been in many ways expected already in the mid
1990s given our knowledge of the vastly disparate parameter space coverages of
X-ray and radio surveys (see \S~\ref{discovery}).

With these recent discoveries in mind, it is time to once again take
stock of the broadband spectral properties of all blazars. Here we
utilize updated identifications from DXRBS (most of which were presented 
by Landt et al. 2001), which result in a much larger sample, combined with a
sample of blazars we have extracted from RGB \citep{lau96,lau98}, to
attack this problem. A preliminary version of some of these findings
was presented previously by \citet{per00}.

The samples are briefly described in \S~2. In \S~3 we comment on the failure
of previous surveys to find X-ray strong FSRQ. In \S~4 we study the SEDs of
our sources and discuss the use of the available data to obtain the peak
frequency of synchrotron emission and other characteristics of blazars. We
discuss the so-called ``blazar sequence'' scenario in \S~5 and further
constraints on blazar emission in \S~6, while \S~7 summarizes our
conclusions. Throughout this paper spectral indices are written $S_{\nu}
\propto \nu^{-\alpha}$ and for consistency with previous work the values $H_0
= 50$ km s$^{-1}$ Mpc$^{-1}$, $\Omega_{\rm M} = 0$, and $\Omega_{\rm \Lambda}
= 0$ have been adopted. We note that our correlations (or lack of) are
basically unchanged for a cosmological model with $H_0 = 65$ km s$^{-1}$
Mpc$^{-1}$, $\Omega_{\rm M} = 0.3$, and $\Omega_{\rm \Lambda} = 0.7$.

\section{The Samples}\label{samples}

\subsection{The DXRBS Blazar Sample}\label{samp_dxrbs} 

DXRBS is the result of correlating the ROSAT WGACAT database \citep{whi95}
with several
publicly available radio catalogs (GB6 and PMN at 5 GHz, NORTH20CM at 1.4
GHz), restricting the candidate list to serendipitous flat-spectrum radio
sources ($\alpha_{\rm r} \le 0.70$). Additionally, a snapshot survey with the
Australia Telescope Compact Array (ATCA) was conducted for $\delta \la
0^{\circ}$ to get radio spectral indices unaffected by variability (and
arc-second positions for the sources south of $\delta = -40^{\circ}$, the
limit of the NRAO-VLA Sky Survey [NVSS]; Condon et al. 1998). The DXRBS X-ray
flux limits depend on the exposure time and the distance from the center of
the ROSAT Position Sensitive Proportional Counter (PSPC) but vary between
$\sim 10^{-14}$ and $\sim 10^{-11}$ erg cm$^{-2}$ s$^{-1}$. Radio 5 GHz fluxes
reach down to $\sim 50$ mJy.

First results from DXRBS were presented in \citet{per98}, while updated
identifications were presented in \citet{lan01}. At the time of writing
(November 2002), DXRBS includes 244 blazars: 200 FSRQ (defined as broad-line
sources with $\alpha_{\rm r} \le 0.50$) and 44 BL Lacs. 
Details on our BL Lac classification are given in \citet{lan01}. Note that
\citet{gio02b} have pointed out that sources with optical spectrum typical of a
galaxy but nuclear SED typical of a blazar should probably be classified as
low-luminosity BL Lacs. Some of our radio galaxies might fall in that category. 
We will address this point in future papers.   

We include in this work only sources belonging to the complete sample, which
fulfills all our selection criteria (see details in Landt et al. 2001). This
is necessary because we want to compare in detail the properties (mainly the
SED) of DXRBS FSRQ and BL Lacs and that requires well-defined samples, not to
introduce any bias. Since we need the sky coverage to de-convolve the observed
distributions, we have excluded sources with PSPC center offsets in the range
$13^{\prime} - 24^{\prime}$, where the sky coverage is difficult to determine
because of the effects of the spacecraft wobble and the rib structure. We are
then left with 134 FSRQ and 31 BL Lacs, for a total of 165 blazars. Only about
10 more sources ($\sim 5\%$ of the total) with $\alpha_{\rm r} \le 0.50$
remain to be identified. The current blazar sample is therefore highly
representative.

\subsection{The RGB Blazar Sample}\label{samp_rgb} 

\citet{lau97} have described in depth the procedures undertaken in
constructing the ROSAT All Sky Survey (RASS)-Green Bank (RGB) catalog of radio
and X-ray emitting sources. Here we review only the essential points. 

The catalog was constructed by correlating the RASS with a radio catalog based
on the 1987 Green Bank 5 GHz survey maps \citep{gre91,gre96}. VLA observations
of all the 2,127 radio/X-ray matches within 100$^{\prime\prime}$ were obtained
to derive arc-second positions (Laurent-Muehleisen et al. 1997).
For the 1,567 sources with radio/X-ray offset $< 40^{\prime\prime}$ optical
counterparts were identified using the Automatic Plate Measuring (APM) scans
\citep{irw94}, which give O (blue) and E (red) magnitudes. \citet{lau97,lau98}
give many new object identifications based upon spectra taken at
various optical telescopes. The identifications published by
\citet{lau97,lau98} include a comprehensive list of BL Lacs, as well as
many radio-loud, broad-line objects (i.e., radio-loud quasars);
however, since radio spectral index was not included in the search
criteria for RGB no list of FSRQ was compiled there.

Some indication that X-ray strong FSRQ were being found in the RGB
sample could be seen from the distribution of X-ray to radio flux
ratios of the newly identified objects \citep{lau98}.  We therefore
decided to use the information in the RGB along with other public
surveys to create a list of all blazars (BL Lacs and FSRQ) in the full
RGB sample. We used Tables 3.3 and 3.4 of \citet{lau96} and Table 3 of
\citet{lau98} to search for both BL Lacs and FSRQ as follows. First,
the RGB sample was cross-correlated with the AGN catalog of
\citet{pad97} to find previously known blazars. Then, we used the
identifications given by \citet{lau98} to include newly identified
sources (note that these are limited to objects with $O \le
18.5$). This resulted in a total of more than 600 classified RGB
sources. Being based on the RASS, the RGB covers a very different
X-ray flux range: X-ray fluxes range between $\sim 2 \times 10^{-13}$
and $\sim 10^{-10}$ erg cm$^{-2}$ s$^{-1}$. Radio 5 GHz fluxes reach down to
$\sim 25$ mJy, roughly a factor of 2 fainter than DXRBS at similar
declinations.

Radio spectral indices, necessary to determine if a broad-line radio source is
an FSRQ or not, were derived by cross-correlating the RGB with the NVSS. For
consistency with DXRBS, we used 5 GHz fluxes from the GB6 survey (and not from
the dedicated VLA observations) for the RGB sources. In doing the
cross-correlation, then, the 1.4 GHz flux from all NVSS sources within a 
3$^{\prime}$
radius (corresponding roughly to the beam size of the GB6 survey) was summed
up (we have already described the rationale connected with doing
this in Perlman et al. 1998). We then isolated a sample of FSRQ ($\alpha_{\rm
r} \le 0.5$).

We checked that the spectral indices derived from the GB6/NVSS radio data were
reliable in the following way. Three hundred and forty-five radio-loud RGB
sources had a value of the radio spectral index in the AGN catalog of
\citet{pad97}, generally obtained from non-simultaneous, single-dish
measurements around a few GHz. There is a very strong ($P > 99.99\%$) linear
correlation with a slope $\sim 0.9$ between the two spectral indices, with a
mean difference $\Delta \alpha = 0.08$ over a range of $\sim 2.5$. This shows
that the method we used to estimate the radio spectral index for the RGB
sample is quite robust.

The RGB blazar sample thus includes 362 blazars: 233 FSRQ and 129 BL
Lacs. To our knowledge the RGB sample is $\sim 44\%$ identified. This fraction goes
up to $\sim 76\%$ for the part of the sample with $O\le 18.5$. Considering
only the sources with $\alpha_{\rm r} \le 0.5$, these fractions do not change
much, being $49\%$ and $73\%$ respectively.
 
\subsection{The Joint Sample}\label{sam_joint} 
 
Overall, the two samples used in this paper include 497 distinct blazars, 342
FSRQ and 155 BL Lacs (30 objects are in common). In terms of its range of
properties, size, depth, and selection criteria this ensemble of objects
represents a unique sample with which to address some of the open questions of
blazar research. 

As an initial step towards studying the broad-band properties of our sources,
we first derive their $\alpha_{\rm ox}$, $\alpha_{\rm ro}$, and $\alpha_{\rm
rx}$ values. These are the usual rest-frame effective spectral indices defined
between 5 GHz, 5,000 \AA, and 1 keV. X-ray and optical fluxes have been
corrected for Galactic absorption. The effective spectral indices have been
k-corrected using the appropriate radio, optical, and X-ray spectral
indices. X-ray spectral indices are available for all DXRBS sources from
hardness ratios, as described in \citet{lan01}. For the RGB objects we
obtained X-ray spectral indices from hardness ratios for $73\%$ and $89\%$ of
FSRQ and BL Lacs respectively from \citet{bri97}, \citet{lau99},
\citet{rei00}, and WGACAT. In the remaining cases we assumed $\alpha_{\rm x} =
1.2$ following \citet{lau98}. Values for the optical spectral indices were
assumed to be the typical ones for the various sub-classes. Finally, optical
magnitudes for broad-line sources were also corrected for the presence of
emission lines according to the prescription of \citet{nat98}.

It is worth mentioning that, unlike DXRBS, all new RGB identifications have $O
\le 18.5$. This has important selection biases associated with it 
(\S \S~\ref{discovery},\ref{SED}). 
In particular, an optical magnitude limit has the effect of imposing an
artificial boundary on the regions of parameter space to which a survey is
sensitive. Namely, only sources with $\alpha_{\rm ro} < \alpha_{\rm ro}$(lim)
and $\alpha_{\rm ox} > \alpha_{\rm ox}$(lim), where these limiting values
depend on radio and X-ray flux, will be included. At the survey radio limit, for
example, only sources with $\alpha_{\rm ro} \la 0.4$ will be included. 
We will discuss the effect of this limitation in \S~\ref{SED}. 

\section{The Discovery of X-ray Strong FSRQ}\label{discovery}

The findings of \citet{pad97}, \citet{per98}, and \citet{pad02}, make it
clear that detecting X-ray strong FSRQ requires large and/or deep
samples. It was already known in 1995 that radio-selected samples find more
frequently objects with broad-band spectra peaking in the IR/optical. The
oft-repeated separation of HBL and LBL on the $\alpha_{\rm
ox},\alpha_{\rm ro}$ plane (e.g., Padovani \& Giommi 1995b), is a direct
by-product of these disjoint survey methods. 

\centerline{\includegraphics[width=9.0cm]{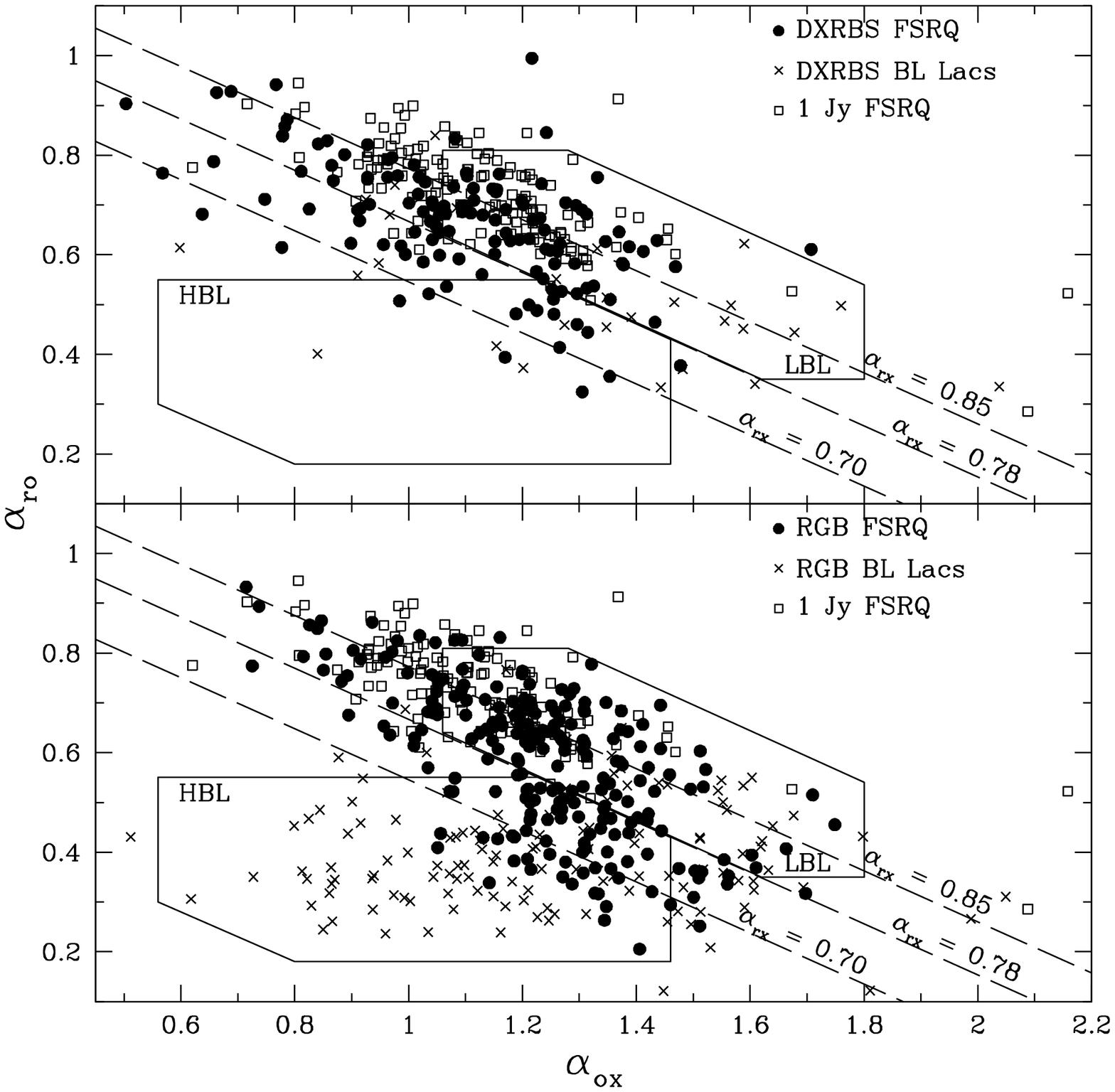}}
\figcaption{The $\alpha_{\rm ro}, \alpha_{\rm ox}$ plane for the DXRBS (top)
and RGB (bottom) samples. Effective spectral indices are defined in
the usual way and calculated between the rest-frame frequencies of 5
GHz, 5000 \AA, and 1 keV. Filled circles represent the DXRBS and RGB
FSRQ, open squares represent the 1 Jy FSRQ, while crosses are the
DXRBS and RGB BL Lacs. The dashed lines represent, from top to bottom,
the loci of $\alpha_{\rm rx} = 0.85$, typical of 1 Jy FSRQ and LBL,
$\alpha_{\rm rx} = 0.78$, the dividing line between HBL and LBL, and
$\alpha_{\rm rx} = 0.70$, typical of RGB BL Lacs. The regions in the
plane within $2\sigma$ from the mean $\alpha_{\rm ro}$, $\alpha_{\rm
ox}$, and $\alpha_{\rm rx}$ values of LBL and HBL are indicated by the
solid lines and marked accordingly.\label{aroaox}}
\centerline{}
\vskip .2in

With this observation in mind, one might expect a similar selection effect to
have been present for FSRQ. Yet up until the mid-1990s, no X-ray survey had
looked for FSRQ, even though re-examinations of the {\it Einstein} EMSS and
Slew Survey databases have revealed samples of FSRQ in both
\citep{per00,per01,wol01}. The only flux-limited samples of FSRQ which
existed prior to that time, in fact, were ``classical'' high flux limit,
radio-selected samples, such as the 1 Jy, produced by surveys which also
contained few, if any HBL (e.g., the 1 Jy sample has only 2 HBL out of
34 BL Lacs total, Padovani \& Giommi 1995b). Thus it was perhaps not a
surprise that these radio-selected samples contained very few X-ray strong FSRQ.

Figure \ref{aroaox} shows the distribution of our sources in the $\alpha_{\rm
ox},\alpha_{\rm ro}$ plane. The figure shows also the 1 Jy FSRQ, which occupy
a region of $\alpha_{\rm ox},\alpha_{\rm ro}$ parameter space with
$\alpha_{\rm rx}$ similar to that typical of LBL (marked in the figure). FSRQ
with low $\alpha_{\rm rx}$ ($\la 0.78$, roughly equivalent to the HBL/LBL
division (although see \S~\ref{SED}), or $L_{\rm x}/L_{\rm r} \ga 10^{-6}$)
constitute only $\sim 5\%$ of the 1 Jy sources with X-ray data. Importantly,
none of the 1 Jy FSRQ fall in the region of the plane within $2\sigma$ from
the mean $\alpha_{\rm ro}$, $\alpha_{\rm ox}$, and $\alpha_{\rm rx}$ values of
HBL, the ``HBL box'', derived by using all HBL in the multi-frequency AGN
catalog of \citet{pad97}\footnote{X-ray data are available for $\sim 65\%$ of
1 Jy FSRQ. The sources without X-ray data have $\alpha_{\rm ro} \ge 0.63$ and
therefore are all outside of the ``HBL box''.}. The fainter radio/X-ray
selected DXRBS and RGB FSRQ, on the other hand, reach much lower values of
$\alpha_{\rm rx}$ and many sources ``invade'' the HBL region.  Indeed, as can
be seen by comparing the two panels of Figure \ref{aroaox}, there is a
progression of $\alpha_{\rm ox},\alpha_{\rm ro}$ from 1 Jy to DXRBS to RGB,
with the exception of the half of the HBL box to the left of the
diagonal line that extends from (0.8,0.55) to (1.4,0.2). We will return to
this subject in \S~\ref{SED}.

\centerline{\includegraphics[width=9.0cm]{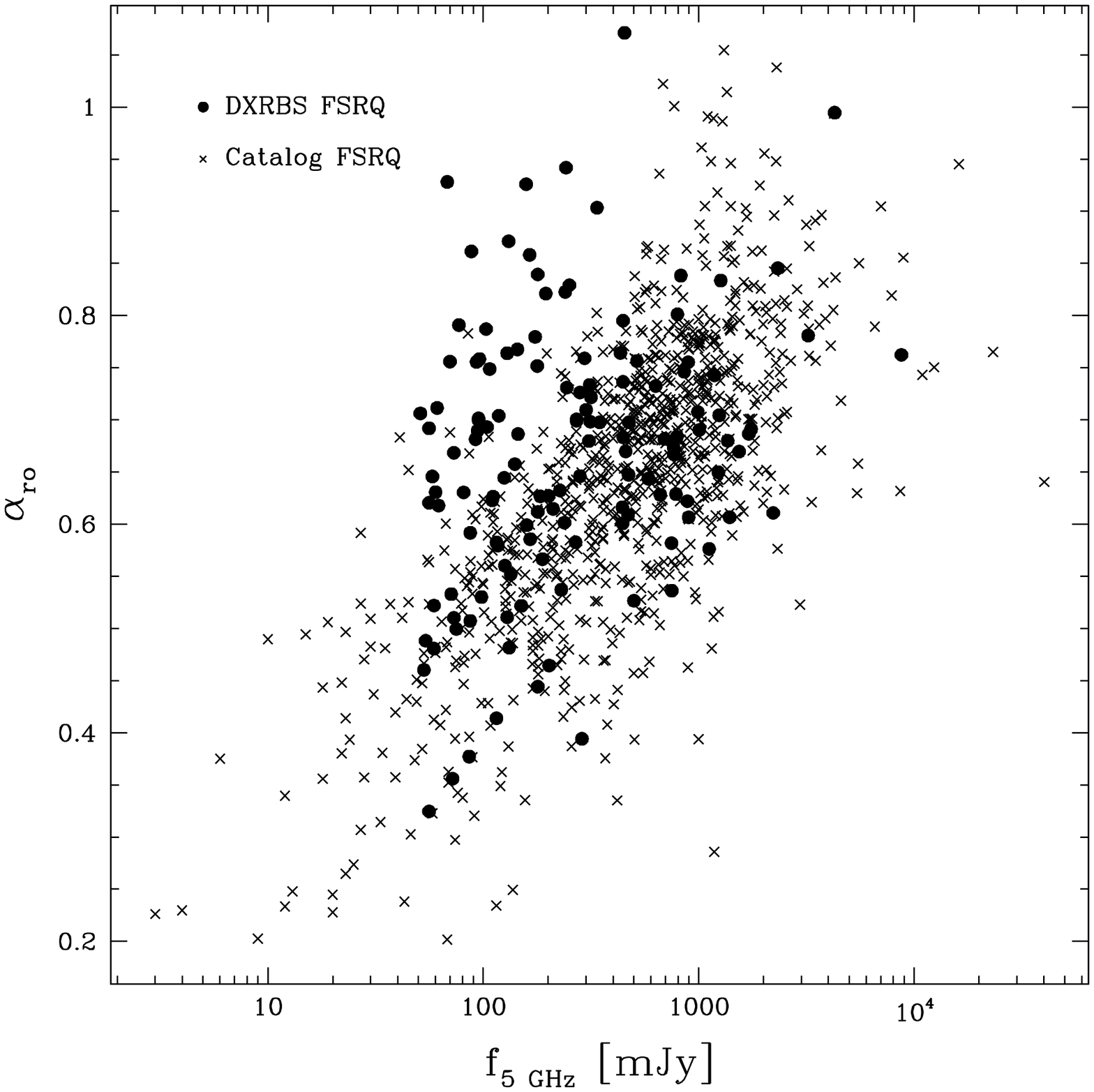}} 
\figcaption{The effective radio-optical spectral index $\alpha_{\rm
ro}$ vs. the 5 GHz radio flux for about 900 FSRQ from the
multi-frequency catalog of \citet{pad97} (crosses) and the DXRBS FSRQ
discussed in this paper (filled points).\label{f6aro}}
\centerline{}
\vskip .2in

Another way of looking at selection effects is to consider Fig. \ref{f6aro},
which plots $\alpha_{\rm ro}$ vs. the 5 GHz radio flux for about 900 FSRQ from
the comprehensive multi-frequency catalog of \citet{pad97}. The observed trend
between radio flux and optical-radio spectral index, with stronger radio
sources having steeper $\alpha_{\rm ro}$ values, is likely due to two separate
effects: 1. the relatively bright optical flux limits of the various samples
which were included in the catalog, responsible for the lack of sources in the
upper left part of the diagram. Indeed DXRBS FSRQ, which reach $V \ga 23$,
start to fill this region (filled points in Fig. \ref{f6aro}); 2. the fact
that well-known, ``classical'' FSRQ, all relatively strong radio sources (3C
273 and 3C 279, for example, have 5 GHz radio fluxes $\sim 16$ and 40 Jy
respectively) do have steep $\alpha_{\rm ro}$, which accounts for the lack of
sources in the lower right part of the diagram. Since HFSRQ need to have
$\alpha_{\rm ro} \la 0.5$ (see Fig. \ref{aroaox}) they will start to be
revealed only at fainter ($\la 200-300$ mJy) radio fluxes, and it is only by
conducting relatively faint radio surveys with X-ray information, like ours,
that these sources can be identified in reasonable numbers.

There could be two possible reasons for the observations pointed out in the
previous two paragraphs. Either some selection effect is at work, perhaps
induced by a correlation between the fraction of HFSRQ and flux, with HFSRQ
being intrinsically more numerous at fainter fluxes/powers, or these sources
are intrinsically rare and their detection in sizeable numbers requires large
samples. We will address this issue in future DXRBS papers. Note that similar
questions are perfectly appropriate as well for BL Lacs and these are the
subject of some debate (see, for example, Fossati 2001, Giommi et
al. 2002c). 

\section{SEDs and the Nature of Blazar Emission}\label{SED}

In the previous section we have shown that recent, relatively deep, blazar
samples with both radio and X-ray information have detected the hitherto
unknown class of FSRQ with effective spectral indices typical of HBL. Based on
our knowledge of BL Lacs, we expect these sources to have a synchrotron peak
frequency $\nu_{\rm peak}$ in the UV/X-ray band. Similarly, we would expect
that the ``intermediate'' BL Lacs in our sample should have synchrotron peak
frequencies intermediate between those seen for HBL and LBL, as demonstrated
for S5 0716+714 by \citet{gio99a}. We then need to study the SEDs of our
sources.

\subsection{$\alpha_{\rm rx}$ Distributions}\label{dist_arx}

The HBL/LBL division is generally done in terms of the X-ray-to-radio flux
ratio or effective radio--X-ray spectral index $\alpha_{\rm rx}$
\citep{pad95b}. While this parameter is broadly related to $\nu_{\rm peak}$
(Padovani \& Giommi 1996; Fossati et al. 1998; but see \S~\ref{arx_peak}), it is 
much easier to derive. Fig. \ref{aroaox} shows that the DXRBS FSRQ {\it and}
BL Lacs seem to cluster at the edge of the HBL region, while the RGB FSRQ and
BL Lacs are moving in progressively. This is due at least partly to the 
different
flux limits of the two surveys, which sample different regions of parameter
space. It does appear, however, that while the RGB BL Lacs populate a much
larger area of the HBL region than DXRBS BL Lacs, this is not the
case for FSRQ. Inclusion of the unclassified RGB sources with optical
information does not alter this conclusion (see also below). Optically fainter
sources will have relatively flat $\alpha_{\rm ox}$ but most importantly
relatively steep $\alpha_{\rm ro}$ and so will be out of the HBL box.

Fig. \ref{arx} shows the fractional $\alpha_{\rm rx}$ distribution of all
DXRBS and RGB blazars.  It is important to note that due to the serendipitous
nature of DXRBS (and other, previous serendipitous surveys, including the
EMSS), the area in which faint X-ray sources could be detected is smaller than
that for brighter X-ray sources. Thus, in order to compare directly
distributions of parameters related to X-ray flux, we need to ``correct'' by
using the appropriate sky coverage -- a necessary step which is seldom done --
otherwise faint sources would be under-represented. We have already done this
for the redshift distribution in Landt et al. (2001) and we refer the reader
to that paper for further details. $\alpha_{\rm rx}$ is obviously related to
X-ray flux so we have de-convolved its distribution, shown in Fig. \ref{arx}.
We follow this practice for all comparisons involving DXRBS and any other
serendipitous survey.

\centerline{\includegraphics[width=9.0cm]{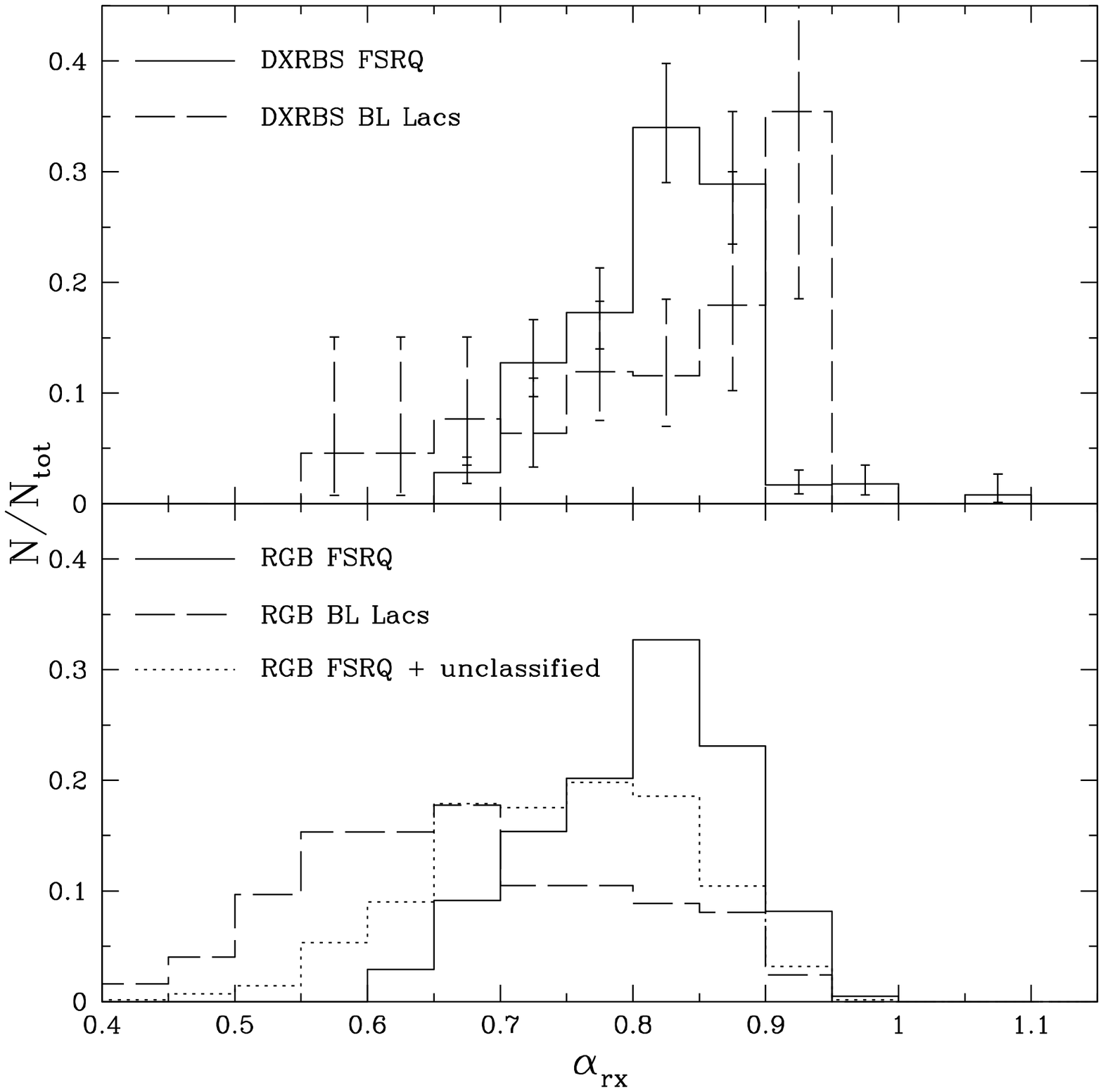}}
\figcaption{The fractional distribution of the X-ray-radio spectral index
$\alpha_{\rm rx}$ for FSRQ (solid lines) and BL Lacs (dashed lines) for the
DXRBS (top) and RGB (bottom) samples. The dotted line represents the
$\alpha_{\rm rx}$ distribution of RGB FSRQ and unclassified radio-loud
sources with $\alpha_{\rm r} \le 0.5$. The DXRBS distributions have been
de-convolved with the appropriate sky coverage. Error bars represent the
$1\sigma$ range based on Poisson statistics. See text for details.\label{arx}}
\centerline{}
\vskip .2in

A few points can be inferred from Fig. \ref{arx}. The $\langle \alpha_{\rm rx}
\rangle$ values for DXRBS FSRQ and BL Lacs are similar, $\sim 0.83$, but the
variances are different at the $99.96\%$ level (Student's t-test). Moreover,
the two distributions are different at the $99.4\%$ level according to a
Kolmogorov-Smirnov (KS) test. The BL Lac $\alpha_{\rm rx}$ distribution, in
fact, is broader than that of FSRQ, with a peak at $\alpha_{\rm rx} \sim 0.9$
and a tail reaching relatively low values ($\sim 0.55$). The skewness of the
BL Lac distribution is negative and equal to $\sim 3.4$ times its standard
deviation, unlike the FSRQ distribution for which the skewness is consistent
with being zero \citep{pre86}. The different nature of the BL Lac and FSRQ
$\alpha_{\rm rx}$ distributions is confirmed by the use of the \kmm
algorithm, developed by \citet{ash94}, which computes for a given distribution
the confidence level at which a single Gaussian fit can be rejected in favor
of a double Gaussian fit. While for the DXRBS FSRQ we find that a single
Gaussian provides a good fit to the $\alpha_{\rm rx}$ distribution, this
possibility is rejected at the $99.9\%$ level for the BL Lacs. The $\sim 10$
still unidentified sources with $\alpha_{\rm r} \le 0.5$ are X-ray faint and
therefore have relatively large $\langle \alpha_{\rm rx} \rangle \sim
0.9$. Based on our identification statistics, we expect most of them to be
FSRQ, thereby somewhat reducing the difference in the $\alpha_{\rm rx}$
distributions at large values in Fig. \ref{arx}.
The fraction of sources with $L_{\rm x}/L_{\rm r} > 10^{-6}$, i.e.,
$\alpha_{\rm rx} \la 0.78$
is $\sim 28\%$ and
$\sim 25\%$ for the BL Lacs and FSRQ respectively. 
More interesting is the
fraction of sources which fall into the region typical of HBL, the ``HBL box''.
This is $\sim
15\%$ and $\sim 9\%$ for the BL Lacs and FSRQ respectively. 
For RGB the $\alpha_{\rm rx}$ distributions for FSRQ and BL
Lacs are also significantly different ($P > 99.99\%$), but in this case the
mean values are also different, with $\langle \alpha_{\rm rx} \rangle \sim
0.80$ and $\sim 0.68$ respectively. The fraction of sources with $L_{\rm
x}/L_{\rm r} > 10^{-6}$ is $\sim 76\%$ and $\sim 35\%$ for the BL Lacs and
FSRQ respectively. The fraction of sources falling into the ``HBL box'' is
$\sim 60\%$ and $\sim 27\%$ for the BL Lacs and FSRQ respectively. While RGB
BL Lacs reach more extreme (i.e., much flatter) $\alpha_{\rm rx}$ values than 
DXRBS ones, that is not the case for FSRQ. 

Could this difference between RGB FSRQ and BL Lacs be
explained as a selection effect? A large fraction ($\sim 60\%$) of the most
X-ray loud RGB BL Lacs ($\alpha_{\rm rx} < 0.7$) were already known because of
dedicated surveys (EMSS, Slew, HEAO-1). Given the fact that the identification
fraction of the flat-spectrum sources in RGB is quite low ($\sim 49\%$) and
that the identified sources in RGB have mostly $O \le 18.5$, it could be that
the most extreme RGB FSRQ are still unclassified. We have then included with
the RGB FSRQ the $\sim 360$ unclassified radio-loud sources with $\alpha_{\rm
r} \le 0.5$ (assuming $z=1$ for the K-correction). Note that this is an
extreme assumption as a substantial fraction of these could be BL Lacs (their
current fraction in the $\alpha_{\rm r} \le 0.5$ sample is $\sim
30\%$). Although the $\alpha_{\rm rx}$ distribution now shifts to lower
values, with $\langle \alpha_{\rm rx} \rangle \sim 0.74$, it is still
significantly different ($P > 99.99\%$) from that of the RGB BL Lacs. The
fraction of sources with $L_{\rm x}/L_{\rm r} > 10^{-6}$ rises from $\sim
35\%$ to $\sim 62\%$, while that which falls into the ``HBL box'' (amongst the
sources with optical information; $\sim 40\%$ of the unclassified sources are
empty fields) rises from $\sim 27\%$ to $\sim 38\%$. As discussed above, this
fraction should probably be considered an upper limit.

We point out that the DXRBS and RGB $\alpha_{\rm rx}$ distributions for both
FSRQ and BL Lacs peak at $\sim 0.7 - 0.8$, that is around the HBL/LBL dividing
line. The previously suggested dichotomy in the BL Lac class then, with most
sources at the two ends of the distribution, was simply due to the selection
effects induced by combining widely disconnected samples selected in different
bands.

We can check what happens at even larger X-ray-to-radio flux ratios by using
two other samples: the EMSS and the Sedentary survey. The EMSS (e.g.,    
Maccacaro et al. 1994) is an X-ray selected sample of sources discovered in
1,435 {\it Einstein} Imaging Proportional Counter (IPC) fields. It reaches
$\sim 5 \times 10^{-14}$ erg cm$^{-2}$ s$^{-1}$ in the $0.3 -3.5$ keV band but
the area covered is a strong function of X-ray flux \citep{gio90}. 
The sample is largely identified (completely so down to $\sim 2 \times
10^{-13}$ erg cm$^{-2}$ s$^{-1}$) and 
comprehensive, dedicated radio observations provide
detections down to $\sim 1$ mJy levels so that quite high values of
X-ray-to-radio flux ratios can be reached. 

Two papers have constructed samples
of FSRQ within the EMSS. Wolter \& Celotti (2001) constructed a sample of EMSS
radio-loud quasars by using a cut $\alpha_{\rm ro} > 0.35$. We have used the
values for the thirteen FSRQ ($\alpha_{\rm r} \le 0.5$) given in their
paper. \citet{per01} did a similar search which yielded those objects plus
two more with $0.2 < \alpha_{\rm ro} < 0.35$ (for consistency with the
commonly used definition of radio-loud sources; these two sources happen to
have the two lowest $\alpha_{\rm rx}$ values). Using these objects, we have
constructed the
$\alpha_{\rm rx}$ distribution of these EMSS FSRQ, again taking into account
the effect of the sky coverage. We have also done the same for the EMSS BL Lac
sample, using the revised list and data of Rector et al. (2000). The results
are as follows: the $\alpha_{\rm rx}$ distributions for FSRQ and BL Lacs are
still significantly different ($P > 99.99\%$), with mean values $\langle
\alpha_{\rm rx} \rangle \sim 0.71$ and $\sim 0.60$ respectively. The minimum
value reached by EMSS FSRQ is $\alpha_{\rm rx} \sim 0.63$, while that of EMSS
BL Lacs is $\sim 0.49$. The fraction of sources with $L_{\rm x}/L_{\rm r} >
10^{-6}$, once the effect of the sky coverage is taken into account, is
$100\%$ and $\sim 78\%$ for the BL Lacs and FSRQ respectively. The fraction of
sources which fall into the ``HBL box'' 
is the same, namely $100\%$ and $\sim 79\%$ for the BL Lacs and FSRQ
respectively. So, although the $\alpha_{\rm rx}$ gap between the two classes
gets narrower, the fact remains that FSRQ do not reach the same values as BL
Lacs.

An even more extreme sample is the Sedentary survey \citep{gio99b}, an
X-ray/radio selected sample based on the RASS Bright Source Catalog (RASSBSC)
and the NVSS, which has a cut at $\alpha_{\rm rx} \la 0.56$. The current
sample is $\sim 90\%$ identified so our conclusions should be relatively
stable. The number of radio-loud broad-lined sources in the sample is
currently around 20 (or $\sim 12\%$), but most of these sources are very close
to the radio-loud/radio-quiet dividing line, both in terms of their
$\alpha_{\rm ro}$ values and their radio powers. Some of these sources could
therefore have their X-ray emission dominated by thermal processes and would
not be the counterparts of HBL. We will discuss the presence of HFSRQ in the
Sedentary survey in a future paper (Giommi et al., in preparation). Table
\ref{tab1} summarizes the results of these comparisons, in terms of mean
$\alpha_{\rm rx}$ values and fractions of HBL and HFSRQ in the DXRBS, RGB, and
EMSS samples.

\begin{table*}
{\footnotesize
\begin{center}
\caption{$\alpha_{\rm rx}$ Statistics. \label{tab1}}
\begin{tabular}{llrlrlrlr}
\tableline\tableline
 &\multispan4{~~~$\langle \alpha_{\rm rx} \rangle$~~}& \multispan2{~~~~~\% $\alpha_{\rm rx} \le 0.78$~~~~~}& \multispan2{~~~~~\% in HBL box~~~~~}\\ 
Sample&FSRQ~~~~~~~~~~~&N&BL Lacs&N&FSRQ~~~~~~~~~~&BL Lacs
&FSRQ~~~~~~~~~~&BL Lacs\\
\tableline
DXRBS & $0.827\pm0.005$ & 134 & $0.84\pm0.02$ &  31 & 25\%& 28\% & 9\% & 15\%\\
RGB   & $0.803\pm0.005$($\ga 0.74$) & 233 & $0.68\pm0.01$ & 129 & 35\% ($\la 62\%$)& 76\%& 27\%($\la 38\%$) & 60\%\\
EMSS  & \phn$0.71\pm0.02$ &  15 & $0.60\pm0.01$ &  41 & 78\% & 100\% & 79\%& 100\%\\
\tableline
\end{tabular}
\end{center} }
\end{table*}

Our results reaffirm the fact that the percentage of HBL/HFSRQ in a sample is
heavily dependent on the survey flux limits or, alternatively, its position on
the X-ray--radio flux plane (see Figs. 1 and 2 of \citet{pad02sax}). Tab. 1
shows very explicitly how these percentages change by moving from DXRBS, a
survey whose limits are in the ``L'' zone, to RGB and EMSS, whose limits move
progressively deeper into the ``H'' zone. 

\subsection{Synchrotron Peak Frequency}\label{peak}

To study in more
detail the synchrotron peak frequencies of our sources we have used the
multi-frequency information at our disposal to extract non-simultaneous SEDs
for our blazars. We restrict ourselves to the DXRBS sample as the
determination of $\nu_{\rm peak}$ requires a lot more effort than $\alpha_{\rm
rx}$ and we did not think it was worth deriving for RGB given its large
incompleteness.

We determined $\nu_{\rm peak}$ for all 165 DXRBS blazars by applying an
homogeneous synchrotron -- inverse self-Compton (SSC) model in the log $\nu$
-- log $\nu f_{\nu}$ plane to the following data:

\begin{enumerate}

\item radio fluxes at two different frequencies (non-simultaneous 1.4 and 5
GHz data for objects with $\delta > -40^\circ$; simultaneous 5 and 8.6 GHz
data for objects with $\delta < -40^\circ$; see Landt et al. 2001);

\item optical flux, in the V band, derived from Palomar Observatory Sky Survey
(POSS I) O and/or E magnitudes as described in \citet{gio99b}; fluxes were
corrected for Galactic absorption and for the presence of emission lines (for
FSRQ) according to the prescription of \citet{nat98};

\item unabsorbed ROSAT X-ray flux at 1 keV, derived from the broad-band
flux and the X-ray spectral index $\alpha_{\rm x}$; 

\item Two Micron All Sky Survey (2MASS) data (if available; Cutri et
al. 2000); 

\item any other NASA/IPAC Extragalactic Database (NED) data. 

\end{enumerate}

We note that here and in the following we deal with the rest-frame peak
frequency, equal to $\nu_{\rm peak}({\rm obs}) \times (1+z)$. A redshift of
0.4 was assumed for the BL Lacs without one.

While we have optical spectra for all objects identified in \citet{per98}
and \citet{lan01}, we did not use those spectra to estimate peak frequencies
for two reasons. First, the DXRBS includes also a significant 
number of previously known sources (about 40\% for the sample used in this 
paper), for which 
optical spectra are not necessarily available in
digital format. We felt it would be inconsistent to use optical spectra
for some objects but not for others. Second, for many objects the spectra
were taken with a slit that was not at parallactic angle due to the
instrumental setups used (see Perlman et al. 1998 and Landt et al. 2001).
This results in a flux loss due to atmospheric differential refraction which
is more pronounced in the blue part of the spectrum and would therefore
significantly affect the optical/UV slope.

The SSC model was been adapted from \citet{tav98} and
assumes that radiation is produced by a population of relativistic electrons
emitting synchrotron radiation in a single zone of a jet that is moving at
relativistic speed and at a small angle to the line of sight. These photons
are subsequently scattered by the same electrons to higher energies via the 
inverse Compton process (note that we ignore Comptonization of external 
photons, which only affects the SED at $\gamma$-ray energies; e.g., Ghisellini 
et al. 1998). The physical parameters that define the model are the jet
radius, the Doppler factor $\delta$, the magnetic field $B$, and four
spectral parameters of the electron population, assumed to follow a power-law
distribution which breaks sharply above a given energy: the normalization, the
two spectral slopes, and the break energy. The Klein-Nishina cross section is
used in the computation of the Compton scattering. 

\begin{figure*}[th]
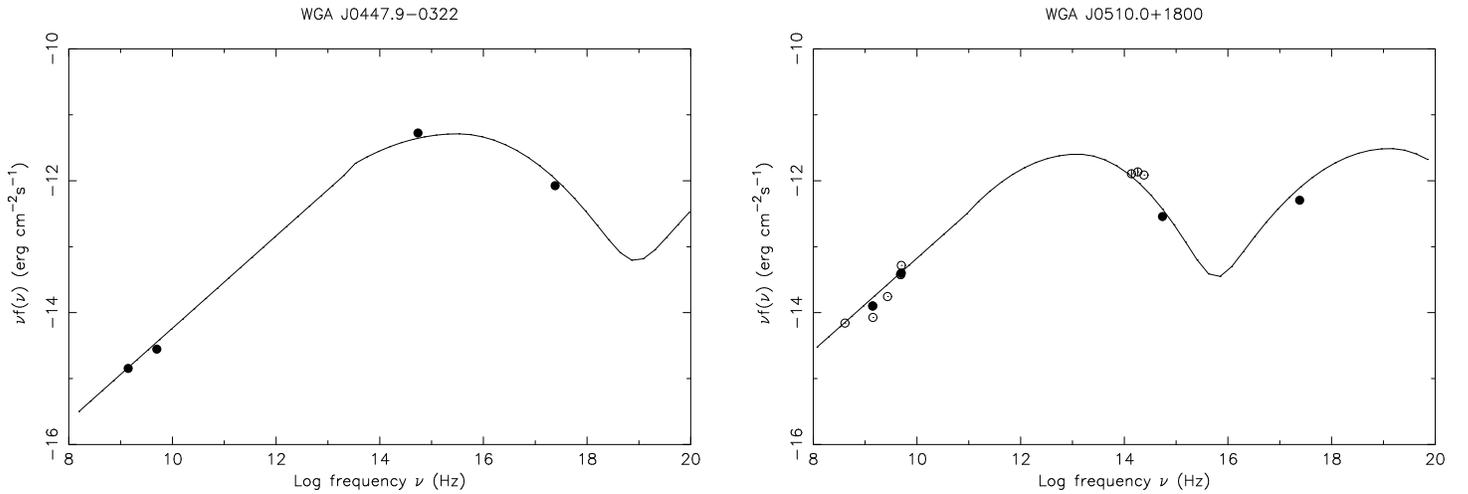

\null
\vskip -5.0truecm
\resizebox{18cm}{!}{\includegraphics[width=9.0cm]{f4a.eps}
\hspace*{0.1cm}\includegraphics[width=9.0cm]{f4b.eps}}
\figcaption{Representative synchrotron self-Compton fits to the
spectral energy distributions of DXRBS FSRQ. Filled points denote
DXRBS data, empty points denote data from NED and 2MASS. The
synchrotron peak frequency for the two sources are $\nu_{\rm peak} =
5.5 \times 10^{15}$ Hz (WGA J0447.9-0322, $z=0.774$, left) and
$\nu_{\rm peak} = 1.7 \times 10^{13}$ Hz (WGA J0510.0+1800, $z=0.416$,
right). See text for details.\label{nupeak_fits_fsrq}}
\end{figure*}

Our aim is to derive a best estimate of $\nu_{\rm peak}$. Given that the SED
of most of our sources is not well sampled we cannot fully constrain the model
parameters. However, the available range was limited by physical
considerations, namely: $\delta \sim 10-20$, $B \sim 0.1 - 5$ G, jet radius
$\approx 10^{-3}$ parsec, and a low-energy slope of the electron distribution
which produces a relatively flat ($< 0.7$) spectral slope in the radio band,
consistent with the sample definition. We have also been guided by the work
done by some of us using the same model to derive $\nu_{\rm peak}$ for a large
sample of bright blazars observed by {\it BeppoSAX} \citep{gio02a} and
from the NVSS-RASS 1 Jy survey \citep{gio02b}. Both samples, in fact,
include many ``classical'' blazars with well sampled SEDs for which the model
parameters were well constrained. In any case, our experience shows that
$\nu_{\rm peak}$, a combination of $B$, $\delta$, and electron break energy,
is relatively well constrained even for relatively large variations of these
parameters, which affect much more the high-energy/inverse Compton part of the
SED (e.g., Massaro et al. 2002).

FSRQ can have a disk (thermal) component in the optical/UV band but this, on
average, makes up only $\sim 15\%$ of the continuum emission in DXRBS FSRQ
\citep{del03}, and therefore should not strongly affect our derivation of
$\nu_{\rm peak}$, at least in a statistical sense, although there are
individual exceptions. This is also consistent with the relatively
steep optical/UV slopes of these objects ($\alpha \sim 1.2$; Landt et
al., in preparation). We think it is also unlikely that the thermal
component can make a substantial contribution in the X-ray band, as
our FSRQ have typical redshifts $\sim 1.5$, which implies that the
ROSAT band corresponds to $0.25 - 6$ keV rest-frame. By comparison,
there is very little evidence for thermal optical/UV emission due to
an accretion disk in BL Lacs, although \citet{cor00} have
suggested that accretion disk illumination is responsible for the
variations in H$\alpha$ luminosity in BL Lacertae itself. In any
case, however, the line luminosity of DXRBS BL Lacs is much
lower than that of FSRQ (Landt et al., in preparation),
so that any contribution to the total optical/UV flux is negligible
for our purposes.

Previous works have derived $\nu_{\rm peak}$ by fitting analytical functions,
such as a parabola or a third-degree polynomial, to the SED of blazars (e.g.,
Sambruna et al. 1996; Fossati et al. 1998). Although the latter alternative
would also allow us to take into account the spectral upturn which is present
when the X-ray flux is dominated by inverse Compton emission, we chose instead
to derive $\nu_{\rm peak}$ using a physical model (SSC) which is widely
accepted as being responsible for the broad-band emission of blazars. We
believe that our approach, although more complex and time consuming, is more
robust especially when dealing with sparsely sampled SEDs, as we are guided by
physics rather than just analytical fitting. We also note that our
multifrequency data are not simultaneous and therefore our fits are also affected
by variability. 

\begin{figure*}[th]
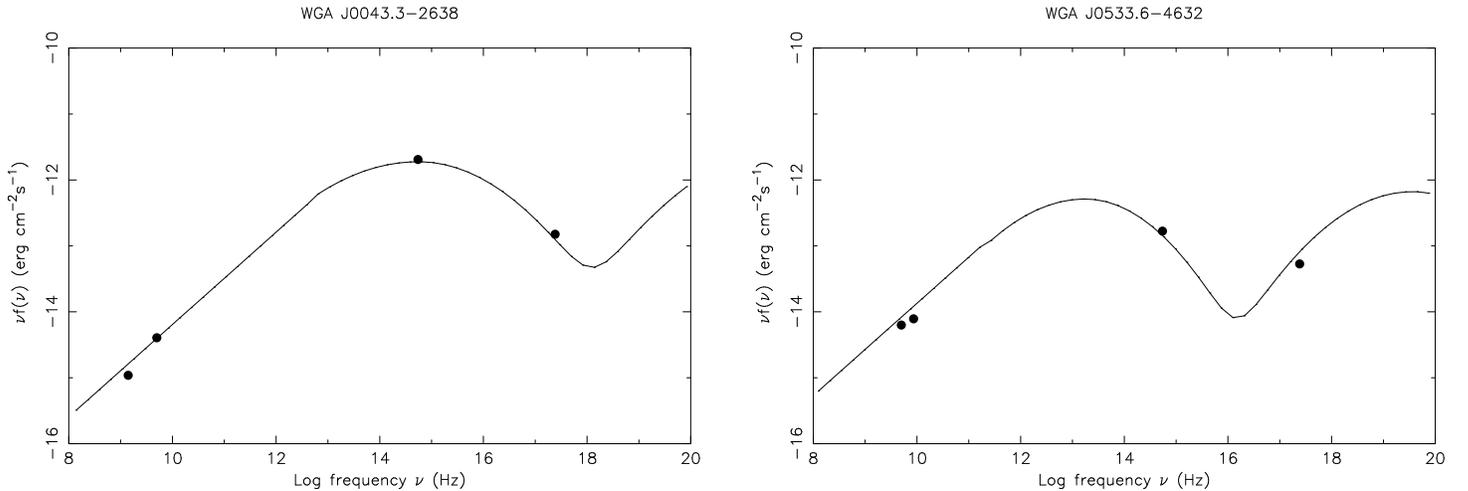

\vskip -5.0truecm
\resizebox{18cm}{!}{\includegraphics[width=9.0cm]{f5a.eps}
\hspace*{0.1cm}\includegraphics[width=9.0cm]{f5b.eps}}
\figcaption{Representative synchrotron self-Compton fits to the spectral energy
distributions of DXRBS BL Lacs. Filled points denote DXRBS data. The
synchrotron peak frequency for the two sources are $\nu_{\rm peak} = 1.1
\times 10^{15}$ Hz (WGA J0043.3-2638, $z=1.002$, left) and $\nu_{\rm peak} =
2.3 \times 10^{13}$ Hz (WGA J0533.6-4632, $z=0.332$, right). See text for
details.\label{nupeak_fits_bllac}}
\end{figure*}

As examples, in Figs. \ref{nupeak_fits_fsrq} and \ref{nupeak_fits_bllac} we
show some representative fits to the SEDs of DXRBS FSRQ and BL Lacs, spanning
a wide range of $\nu_{\rm peak}$. We stress again that the sampling of the
SEDs is scanty, being based in many cases only on a few, non-contemporaneous
data points, so that the precise value of $\nu_{\rm peak}$ for a given source
can be somewhat uncertain. However, we do believe that its {\it statistical}
use is warranted. This is confirmed by plotting the optical/UV slopes, which
are measured for 68 ($\sim 51\%$) of our FSRQ and did not enter in our
derivation of the synchrotron peak frequency, vs. $\nu_{\rm peak}$. A very
strong anti-correlation ($P = 99.96\%$) is observed between the two
quantities, as expected if our $\nu_{\rm peak}$ estimates were indeed
reliable. For relatively small $\nu_{\rm peak}$ values, in fact, one expects
the optical band to sample the steep tail of the synchrotron component, while
when the synchrotron peak moves to progressively larger energies the optical
slope becomes flatter as the optical band samples the lower frequency, flatter
parts of the synchrotron emission.

Figure \ref{nupeak_histo}, which shows the $\nu_{\rm peak}$ distribution of
all our sources, illustrates the fact that indeed, the SEDs of some DXRBS FSRQ
peak in the UV/X-ray band. As in the
case of $\alpha_{\rm rx}$, the $\nu_{\rm peak}$ distributions are also
corrected for the effect of the sky coverage. The FSRQ distribution ranges
between $10^{12}$ and $10^{16}$ Hz, with $\langle \nu_{\rm peak} \rangle =
10^{13.70\pm0.06}$ Hz, while the BL Lac distribution ranges between $6 \times
10^{12}$ and $8 \times 10^{16}$ Hz, with $\langle \nu_{\rm peak} \rangle =
10^{14.1\pm0.1}$ Hz. On average, then, BL Lacs have a synchrotron peak
frequency a factor $\sim 2.5$ larger than that of FSRQ. The two distributions
are also different at the $99.97\%$ level according to a KS test. As was the
case for $\alpha_{\rm rx}$, the BL Lac distribution is broader than the FSRQ
one, with a tail reaching higher values. For example, while $\sim 64\%$ and
$\sim 8\%$ of BL Lacs have $\nu_{\rm peak} > 10^{14}$ and $10^{15}$ Hz, these
fractions go down to $\sim 31\%$ and $\sim 4\%$ for FSRQ. 

We cannot think of any reason why our method should result in lower $\nu_{\rm
peak}$ for FSRQ. Quite the opposite: both the thermal component (which is
small, as discussed above) and the emission lines contamination of the optical
flux (for which we correct anyway) would move the peak to {\it higher} values
(at least with relatively low $\nu_{\rm peak}$).
We therefore regard this difference as real. 

It is likely that this difference in $\nu_{\rm peak}$, actually, is even
larger than shown here. Due to its serendipitous nature, DXRBS is subject to
incompleteness at high fluxes, since bright targets are excluded by definition
not to bias the sample. We can quantify this in the following way. 
Approximately 50 BL Lacs with radio flux above our threshold, mostly from the
1 Jy, Slew, and EMSS samples, and off the Galactic plane have been observed by
ROSAT as targets. This corresponds to a surface density $\approx 2 \times
10^{-3}$ deg$^{-2}$, i.e. roughly $3-4$ sources which
could not be included in our survey. This effect is more severe at high
$\nu_{\rm peak}$ values where only a few objects have been detected (see
Fig. \ref{nupeak_histo}) and the loss of even a small number of sources could
make a major difference. Note in fact that many of these BL Lacs are HBL. As
regards FSRQ, although the numbers are somewhat larger, the effect at high
$\nu_{\rm peak}$ values is zero because none of the previously known (and
therefore likely ROSAT targets) FSRQ had high $\nu_{\rm peak}$.  
 
\centerline{\includegraphics[width=9.0cm]{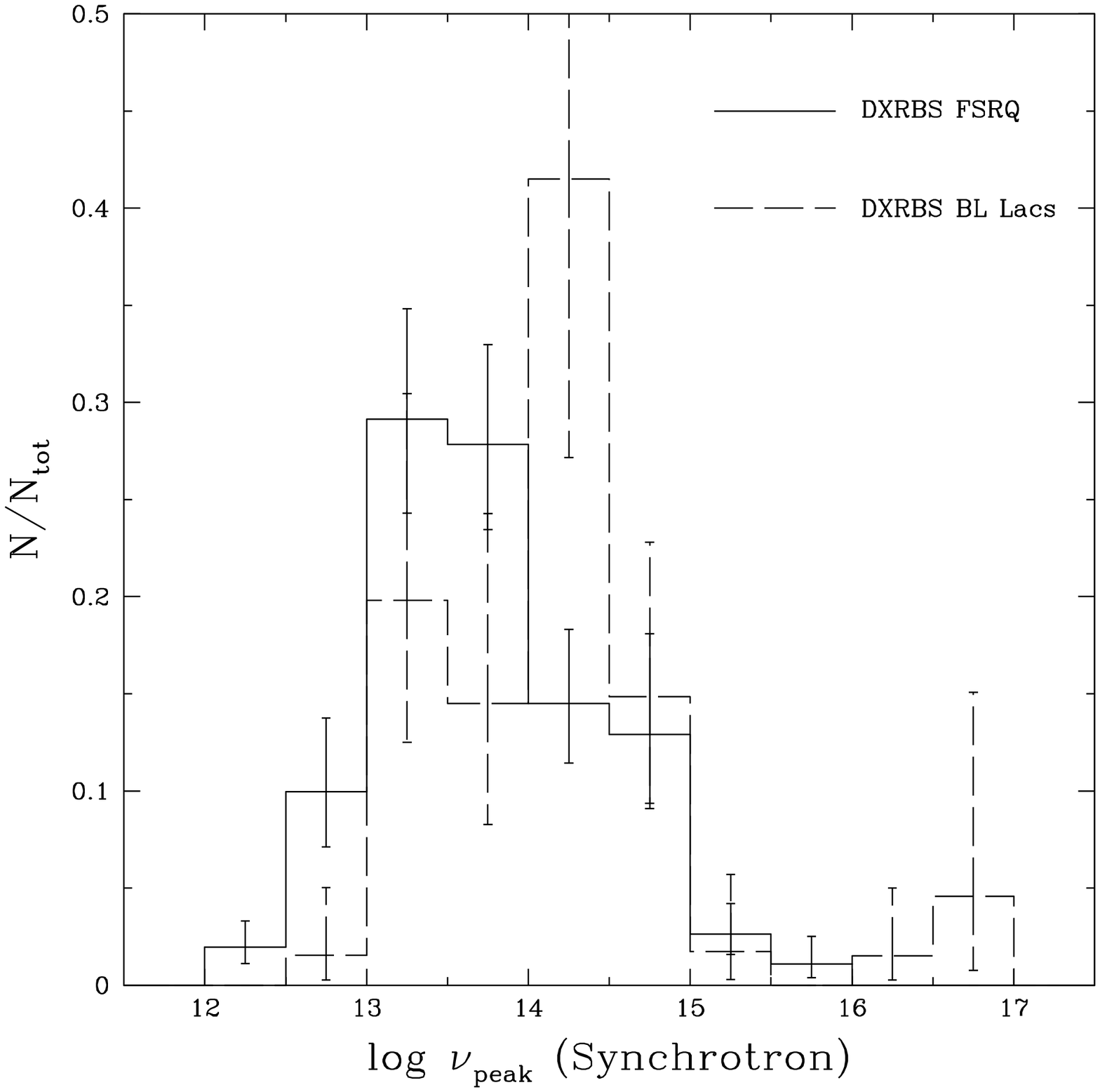}}
\figcaption{The distribution of the synchrotron peak frequency for FSRQ (solid
line) and BL Lacs (dashed line) for the DXRBS sample. The distributions have
been de-convolved with the appropriate sky coverage. Error bars represent the
$1\sigma$ range based on Poisson statistics. See text for
details.\label{nupeak_histo}}
\centerline{}
\vskip .2in

\centerline{\includegraphics[width=9.0cm]{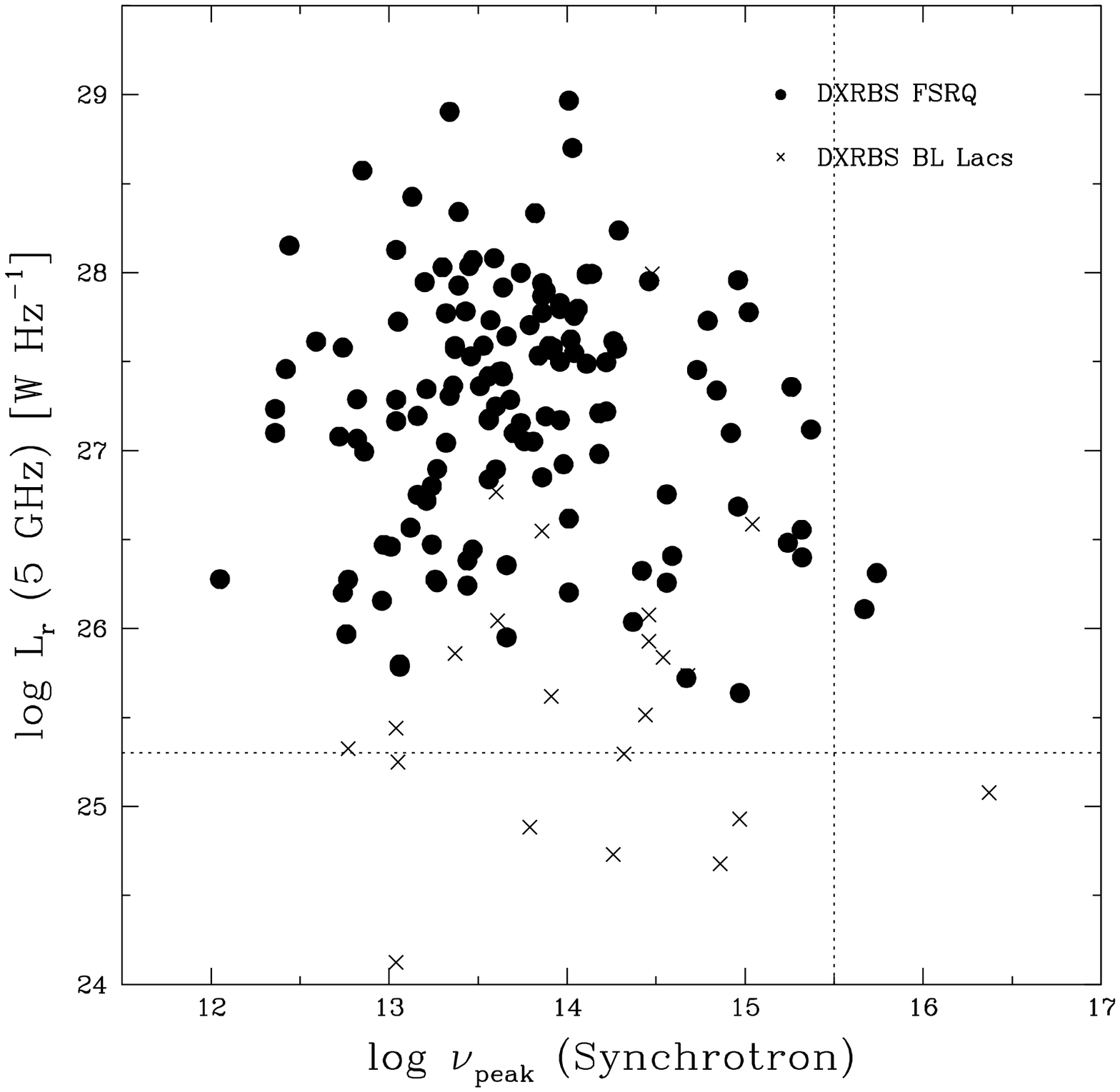}}
\figcaption{Radio power at 5 GHz vs. the synchrotron peak frequency for FSRQ
(filled points) and BL Lacs (crosses) for the DXRBS sample. The dotted lines
denote the two quadrants (top-left and bottom-right) occupied by the sources
studied by \citet{fos98}. See text for details.\label{nubreak_loglr}}
\centerline{}
\vskip .2in

While already in DXRBS then there is evidence that FSRQ do not appear to reach
the more extreme $\alpha_{\rm rx}$ and $\nu_{\rm peak}$ values of BL Lacs,
this tendency is confirmed by studying the RGB and EMSS samples, which sample
the blazar population deeper into the region of lower $\alpha_{\rm rx}$. The
fact that the Sedentary survey, which samples the extreme HBL zone, has found
very few, if any, FSRQ, is also suggesting that there might be an intrinsic
limit to the $\nu_{\rm peak}$ values which can be reached by strong-lined
blazars.

\section{The Blazar Sequence}\label{sequence} 

\citet{fos98} and \citet{ghi98} have proposed that some
blazar properties can be accounted for by an inverse correlation between
intrinsic power and the synchrotron peak frequency, the so-called ``blazar
sequence''. The peak of the emission is related to the electron energy, as
$\nu_{\rm peak} \propto B \gamma_{\rm break,e}^2$, with $\gamma_{\rm break,e}$
a characteristic electron energy which is determined by a competition between
acceleration and cooling processes. Therefore, less powerful sources (where
the energy densities are relatively small) should reach a balance between
cooling and acceleration at larger $\nu_{\rm peak}$, while in more powerful
sources there is more cooling and the balance is reached at smaller $\nu_{\rm
peak}$.

Figure \ref{nubreak_loglr} plots radio power at 5 GHz versus $\nu_{\rm peak}$
for the DXRBS sources. The dotted lines denote the two quadrants (top-left and
bottom-right) occupied by the sources studied by \citet{fos98}, which
belonged to the 1 Jy and Slew BL Lac samples and the 2 Jy FSRQ sample. 

A few points can be made about this figure. First, as already shown in
Fig. \ref{nupeak_histo}, DXRBS BL Lacs reach $\nu_{\rm peak}$ values slightly
higher than DXRBS FSRQ. (Based on the analysis shown in \S~\ref{dist_arx} this
difference is likely to become more pronounced for more X-ray extreme surveys
such as EMSS.) Second, DXRBS sources are starting to occupy regions of this
plot (top-right and particularly bottom-left) which were ``empty'' in the
original plot of \citet{fos98}. DXRBS reaches lower radio powers than the 1 Jy
sample and therefore detects low-power LBL. In particular, out of the 21 BL
Lacs with $\nu_{\rm peak} < 10^{15.5}$ Hz and redshift information, 7 (or
$33\%$) ``invade'' the low-power part ($L_{\rm r} < 10^{25.3}$ W/Hz) of the
plot. Third, no correlation ($P \sim 93\%$) is present between radio power and
$\nu_{\rm peak}$ for the whole sample, or for the FSRQ and BL Lac samples
separately. We also note that the scatter in the plot is very large, reaching
four orders of magnitude in power for $10^{13} \la \nu_{\rm peak} \la 10^{15}$
Hz. Therefore, by using an homogeneous, well-defined sample which includes
both FSRQ and BL Lacs over a relatively wide region of parameter space, the
Fossati et al. correlation is not confirmed. Fourth, an upper envelope,
however, seems to be present in the right part of the diagram. For example, all
sources with $L_{\rm r} > 10^{27.5}$ W Hz$^{-1}$ have $\nu_{\rm peak} \la
10^{15}$ Hz while sources with $L_{\rm r} > 10^{28}$ W Hz$^{-1}$ have
$\nu_{\rm peak} \la 10^{14}$ Hz.

We have checked the radio power and $\nu_{\rm peak}$ values and distributions
for DXRBS blazars in and out of the HBL box. We find that the two classes have
indistinguishable radio powers but significantly different synchrotron peak
frequencies, with mean values $\langle \nu_{\rm peak} \rangle \sim 10^{15.2\pm
0.1}$ and $\sim 10^{13.66\pm 0.05}$ Hz respectively, again in contrast to the
proposed correlation. 

We note that the lack of high $\nu_{\rm peak}$ -- high radio power blazars
could also be due to a selection effect, as discussed by \citet{gio02c} (see
also \citep{pad02}). This combination, in fact, implies a dominance of
non-thermal emission over the host galaxy and emission line components, making
the redshift determination very hard if not outright impossible. Objects of
this kind, which would populate the upper right part of
Fig. \ref{nubreak_loglr}, would then be excluded. We note, however, that only
four DXRBS BL Lacs have no redshift and $\nu_{\rm peak}> 10^{14.5}$ Hz.

\centerline{\includegraphics[width=9.0cm]{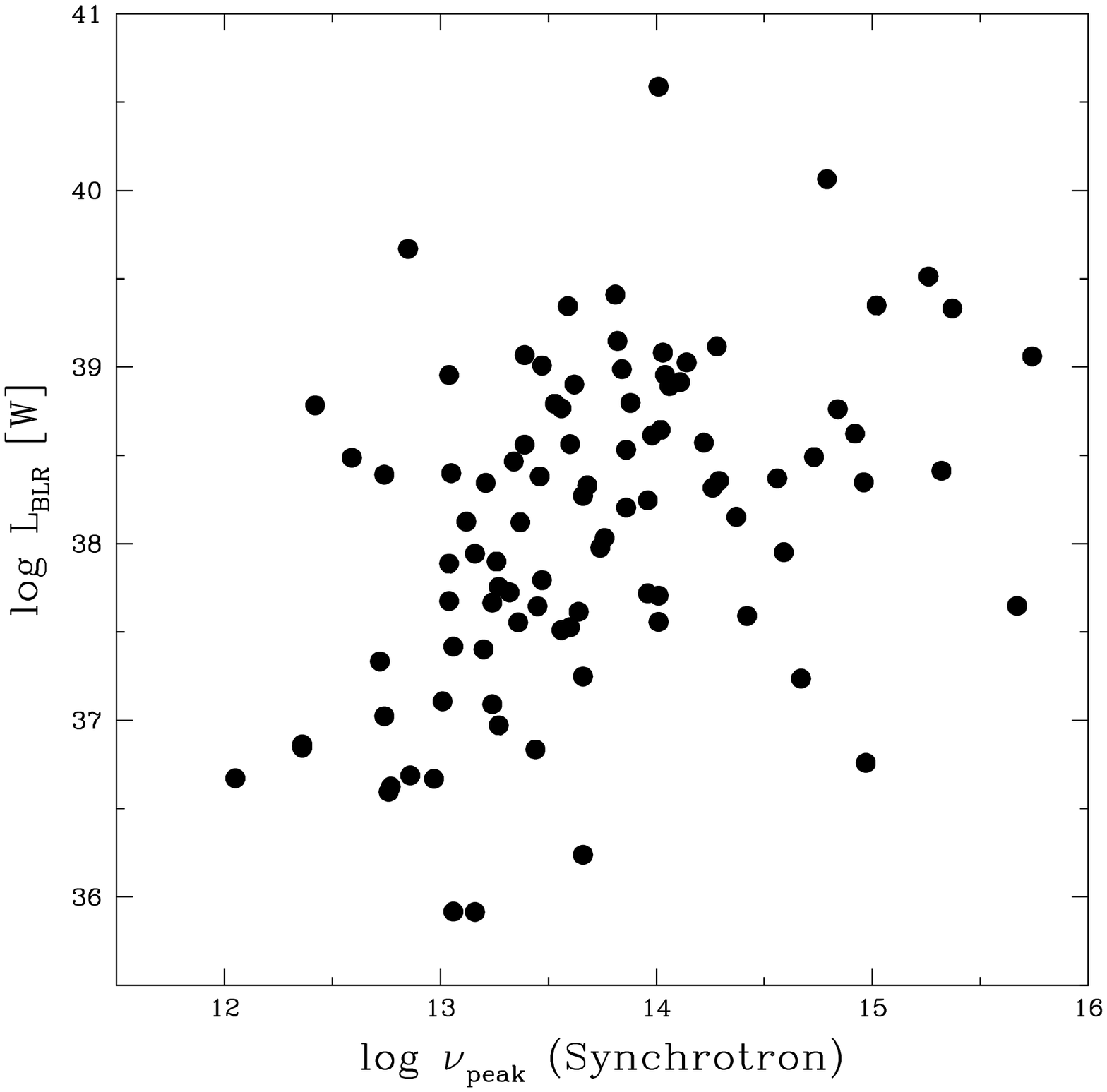}} 
\figcaption{Broad Line Region luminosity vs. the synchrotron peak
frequency for the DXRBS FSRQ. See text for
details.\label{nubreak_loglblr}}
\centerline{}
\vskip -0.39truecm

It is interesting to explore possible correlations between $\nu_{\rm peak}$
and other powers. The correlation suggested by \citet{fos98} and \citet{ghi98}
was between synchrotron peak frequency and {\it intrinsic} power. As our
sources are all flat-spectrum their radio power is strongly affected by
beaming and this could influence the interpretation of Fig.
\ref{nubreak_loglr}. We have then evaluated two intrinsic powers for our FSRQ,
namely the Broad Line Region (BLR) luminosity, $L_{\rm BLR}$, following
\citet{cel97}, and the kinetic jet power, $L_{\rm jet}$, following
\citet{del03}.

$L_{\rm BLR}$ is an isotropic quantity, related to the ionizing, disk
(thermal) emission via the covering factor $f_{\rm cov}$, i.e., $L_{\rm disk}
= f_{\rm cov}^{-1} L_{\rm BLR}$, with $f_{\rm cov} \approx 10\%$ for FSRQ
\citep{del03}. Figure \ref{nubreak_loglblr} plots $L_{\rm BLR}$ vs. $\nu_{\rm
peak}$ for the 94 ($\sim 70\%$) DXRBS FSRQ for which we could find the
relevant BLR line fluxes. 
We do not include here BL Lacs because only 6 of them display the lines needed
to derive $L_{\rm BLR}$ and even in those cases the lines are narrow. No hint
of an inverse correlation between $L_{\rm BLR}$, proportional to disk power,
and $\nu_{\rm peak}$ is present in the figure, the opposite is rather apparent.
There is in
fact a strong correlation ($P > 99.99\%$), with a large scatter, between the
two quantities, with $L_{\rm BLR} \propto \nu_{\rm peak}^{0.51\pm0.11}$. It
could be argued that the stronger the disk emission, the larger its
contribution to the optical/UV flux, and the higher the estimated $\nu_{\rm
peak}$. However, this is not a very likely explanation. First, as mentioned in
\S~\ref{peak}, \citet{del03} have shown that the disk (thermal) component in
DXRBS FSRQ is only $\sim 15\%$ on average. Second, we find no correlation
between $L_{\rm BLR}$ or $\nu_{\rm peak}$ and the ratio of disk to total
emission as defined in \citet{del03}. This would be expected if larger BLR
luminosities and/or larger peak frequencies were due to a stronger disk
component. 

It is more difficult to estimate the total kinetic jet power $L_{\rm jet}$ as
it depends on many uncertain astrophysical parameters. We have derived it
according to the prescriptions of \citet{del03}, which we summarize here
briefly. We have used a theoretical relationship obtained by \citet{wil99} to
link jet power to extended radio emission at 151 MHz (using $c=20$ in eq. 10
of \citet{del03}, which agrees with another independent method), estimating
the latter from the total power at 5 GHz and a correlation between
core-dominance parameter and radio spectral index. We stress that, although
the precise values of $L_{\rm jet}$ for a given source can be somewhat
uncertain, we believe that its {\it statistical} use is warranted.
 
Figure \ref{nubreak_logljet} plots the kinetic jet power vs. $\nu_{\rm peak}$
for our sources. No correlation is present between the two quantities. As
$L_{\rm jet}$ is an intrinsic power, again this is in contrast with the
``blazar sequence'' of \citet{fos98} and \citet{ghi98}. Similarly to Fig.
\ref{nubreak_loglr}, however, in Fig. \ref{nubreak_logljet} there might be
indications of an upper envelope in the right part of the diagram. For
example, sources with $L_{\rm jet} > 10^{39}$ W all have $\nu_{\rm peak} \la
10^{15}$ Hz while sources with $L_{\rm r} > 10^{40}$ W have $\nu_{\rm peak}
\la 10^{14}$ Hz.

\centerline{\includegraphics[width=9.0cm]{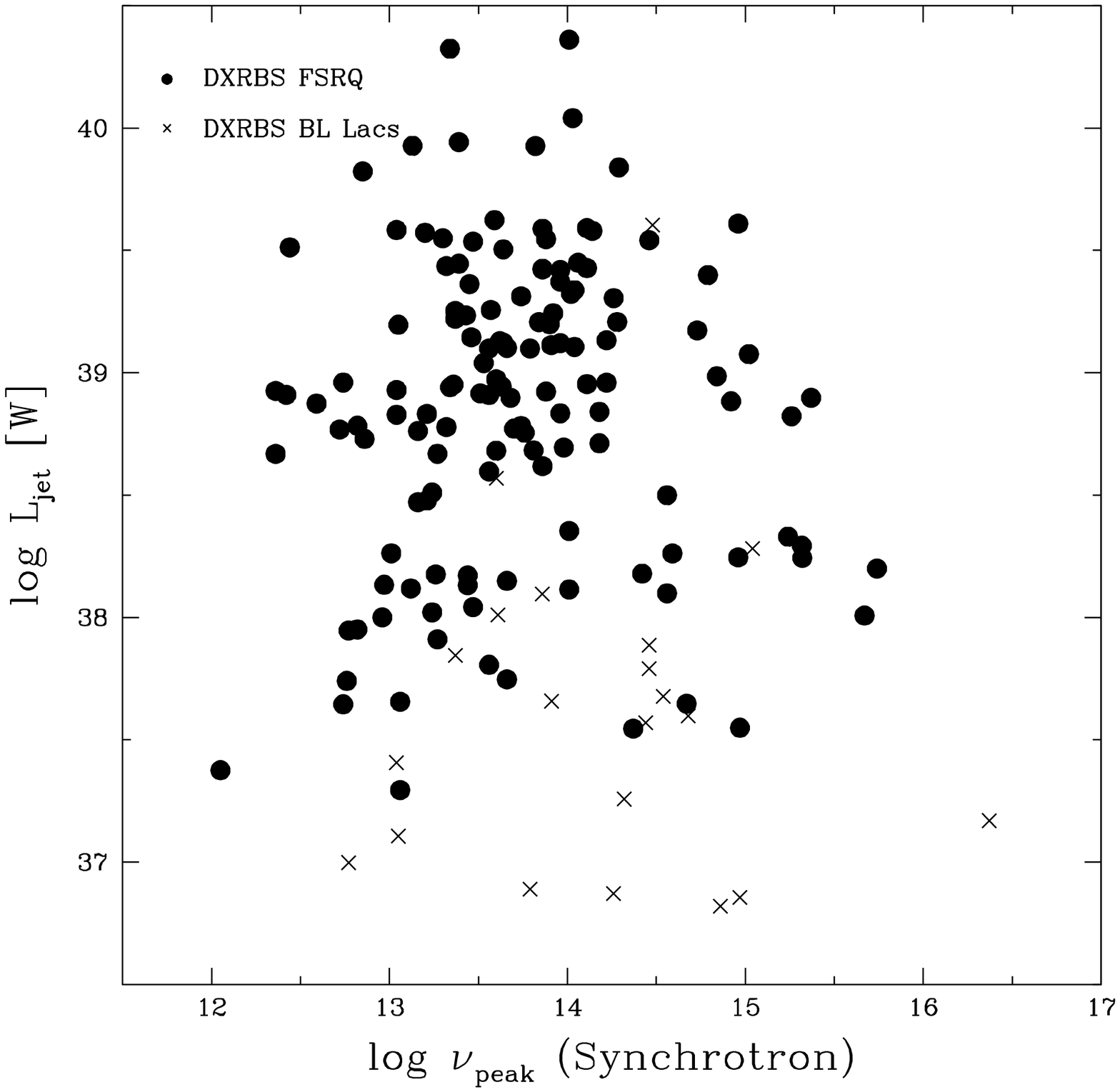}}
\figcaption{Kinetic jet power vs. the synchrotron peak frequency for 
the DXRBS FSRQ (filled points) and BL Lacs (crosses). See text for
details.\label{nubreak_logljet}}
\centerline{}
\vskip .2in

\section{Further Clues}\label{discuss}

We have shown the existence of significant numbers of FSRQ with broad-band
spectra similar to those of HBL, that is with synchrotron peak frequencies
higher than those of ``classical'' FSRQ and reaching the UV band. We now
concentrate on DXRBS to further investigate the implications of our
findings. 

\subsection{X-ray Spectral Slopes}\label{slopes}

The study of the X-ray emission of the FSRQ with synchrotron peak frequencies
in the UV/X-ray band is vital. The X-ray emission of HFSRQ, in fact, should be
synchrotron in nature, in contrast to that of most FSRQ, where it is dominated
by inverse-Compton emission (e.g., Sambruna et al. 1996); a dichotomy similar
to that exhibited by LBL and HBL. In BL Lacs, in fact, observational evidence
points to a different origin for the X-ray emission of HBL and LBL (e.g.,
Padovani \& Giommi 1996). In HBL, the X-ray continuum appears to be an
extension of the synchrotron emission seen at lower energies, consistent with
their steep ($\alpha_{\rm x} \sim 1.5$) X-ray spectra in the ROSAT band. In
LBL, the X-ray continuum is more likely due to inverse Compton emission,
consistent with their harder ($\alpha_{\rm x} \sim 1$) spectra. {\it BeppoSAX}
observations of BL Lacs are confirming this picture \citep{wol98,pad01}.
Indeed, \citet{pad02} have presented evidence for at least one HFSRQ with
relatively high $\nu_{\rm peak} \sim 2 \times 10^{16}$ Hz and steep
($\alpha_{\rm x} \sim 1.5$) synchrotron X-ray spectrum and two possible
``intermediate'' sources with $\nu_{\rm peak} \approx 10^{15}$ Hz. 

We have derived X-ray spectral indices in the $0.4 - 2.0$ keV range for the
DXRBS FSRQ using hardness ratios following \citet{pad96}. This method, when
applied to relatively bright X-ray sources, is relatively robust and compares
well with the results of a proper spectral fit to the full pulse-height
analyzer (PHA) spectrum. One important limitation of this method is the effect
of the PSPC background, which is not subtracted off the WGACAT hardness ratios
and becomes more and more important close to the sensitivity limit, especially
at large off-axis angles. \citet{fio98} have discussed this at length and
shown that for sources with signal-to-noise ratio (SNR) $> 7$ background
contamination is not important. However, many of the \citet{fio98} sources
were targets and therefore at relatively small PSPC offsets. As DXRBS is a
serendipitous survey, we have excluded all targets and therefore, on average,
our sources are at larger PSPC offsets, where the PSPC point spread function
(PSF) degrades significantly, increasing the seriousness of the background
contamination. We have then conservatively chosen a higher SNR cut of 10,
which reduces our sample to 54 sources, 39 FSRQ and 15 BL Lacs. We stress that
these effective spectral indices should still be regarded as an estimate of
the ``average'' X-ray spectral shape and are therefore most suitable for
statistical studies. Errors on these spectral indices ($1\sigma$) were derived
as described in \citet{pad96} and are typically $\sim 0.2$.

\vspace{2truecm} 
\centerline{\includegraphics[width=9.0cm]{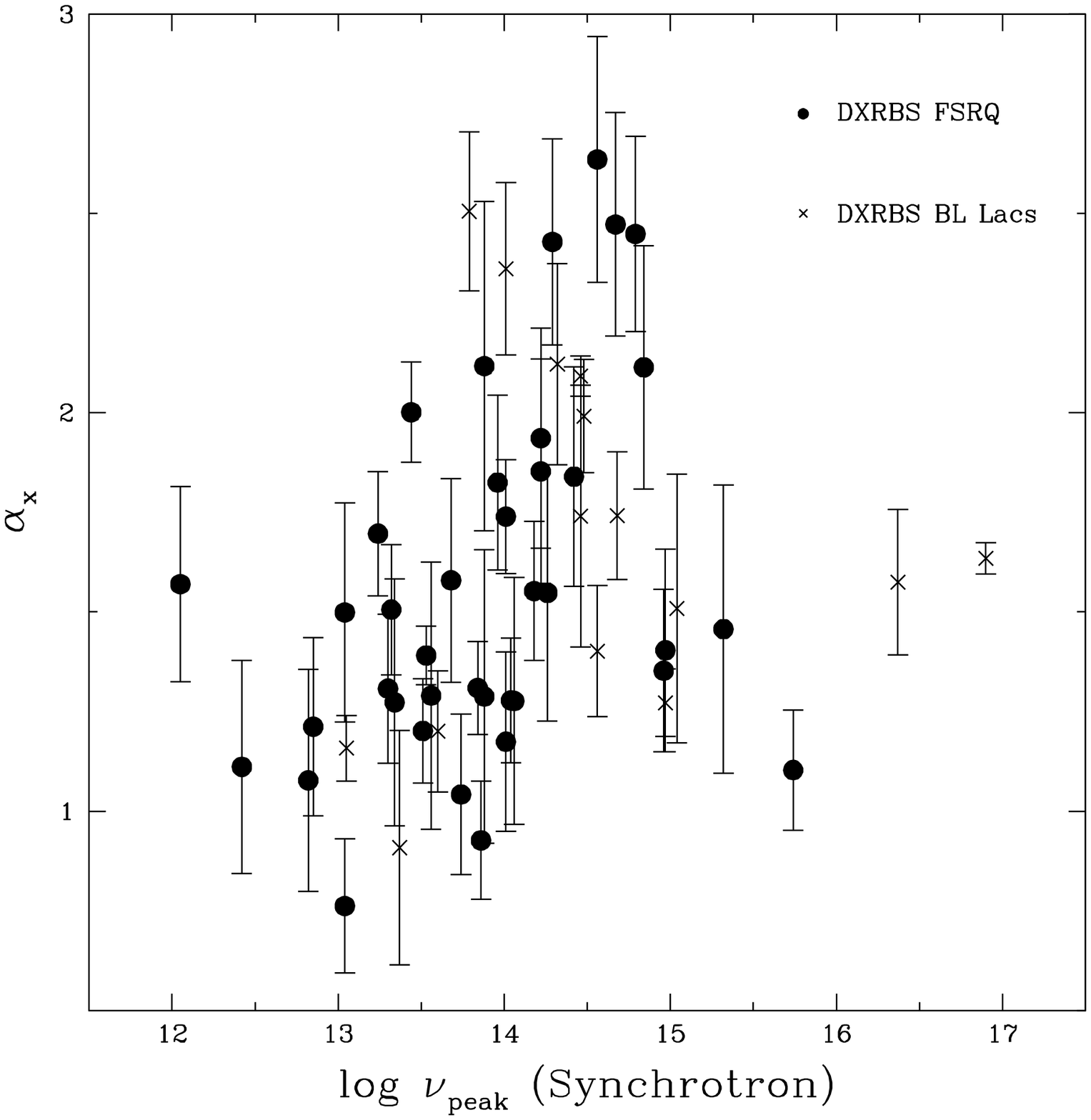}}
\figcaption{ROSAT X-ray spectral index vs. the synchrotron peak frequency for
DXRBS FSRQ (filled points) and BL Lacs (crosses). Error bars represent
1$\sigma$ uncertainties.\label{alpha_nubreak}}
\centerline{}
\vskip .2in

Figure \ref{alpha_nubreak} shows the ROSAT X-ray spectral index vs. the
synchrotron peak frequency for DXRBS FSRQ (filled points) and BL Lacs
(crosses). We note that $\alpha_{\rm x}$ is relatively flat ($\sim 1-1.5$) and
constant for $\nu_{\rm peak} \la 10^{14}$ Hz. For $10^{14} \la \nu_{\rm peak}
\la 10^{15}$ Hz, $\alpha_{\rm x}$ steepens to reach values up to $\sim 2.5$.
Above $\nu_{\rm peak} \sim 10^{15}$ Hz $\alpha_{\rm x}$ flattens again to
reach values $\sim 1-1.5$. Fig. \ref{alpha_nubreak} shows a trend similar to
that displayed by the BL Lacs included in Fig. 6 of \citet{pad96}, despite the
fact that $\sim 70$\% of the sources are FSRQ. We then infer that the
interpretation put forward for BL Lacs applies to our FSRQ as well. Namely, at
low $\nu_{\rm peak}$ values flat inverse Compton emission dominates. For
``intermediate'' values the steep tail of the synchrotron component enters the
ROSAT band. Finally, when $\nu_{\rm peak}$ gets even closer to the X-ray band,
the X-ray spectrum will flatten out again, because the ROSAT band is now
sampling the top of the synchrotron emission.

\centerline{\includegraphics[width=9.0cm]{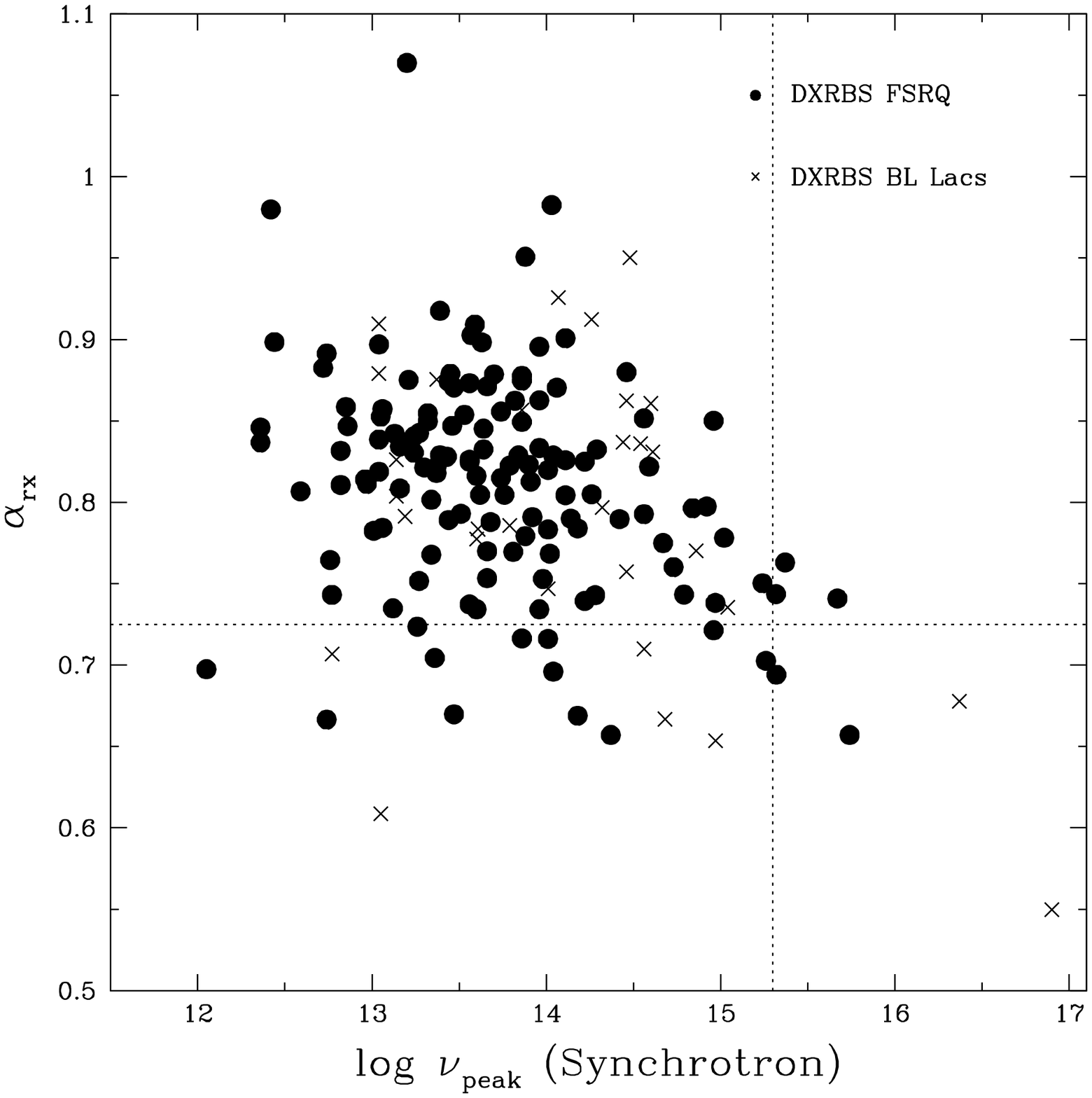}}
\figcaption{Radio-X-ray spectral index $\alpha_{\rm rx}$ vs. synchrotron peak
frequency $\nu_{\rm peak}$ for FSRQ (filled points) and BL Lacs (crosses) for
the DXRBS sample. The dotted lines at $\alpha_{\rm rx} = 0.725$ and $\log
\nu_{\rm peak} = 15.3$ denote the two quadrants (top-left and bottom-right)
occupied by the sources studied by \citet{fos98}. See text for
details.\label{arx_nubreak}}
\centerline{}
\vskip .2in

We note that {\it BeppoSAX} observations of four HFSRQ candidates have given
mixed results: one source is clearly synchrotron dominated, another one is
clearly inverse Compton dominated, while two others have a flat X-ray spectrum
with evidence of steepening at low energies, similar to intermediate BL Lacs
\citep{pad02}. It is, however, worth noting that the synchrotron-dominated
source is the one with the lowest $\alpha_{\rm rx}$ and right in the middle of
the HBL box, while the Compton-dominated object has the largest $\alpha_{\rm
rx}$ and is at the edge of the box. {\it BeppoSAX} spectra of ``classical'' (2
Jy) FSRQ, on the other hand, are all consistently flat, with $\alpha_{\rm x}
\sim 0.7$, perfectly explained by inverse Compton emission
\citep{tav02,gio02a}. \citet{sam00} reported on {\it ASCA} observations of
four FSRQ characterized by steep ROSAT spectra ($\alpha_{\rm x} \sim
1.3$). The sources were all found to have flat hard X--ray spectra, with
$\alpha_{\rm x} \sim 0.8$. Sambruna et al. discuss their results in terms of
relatively high synchrotron peaks and thermal emission extending into the
X--ray band. Importantly, their sources sample a region of parameter space
widely different from ours. Their effective spectral indices place them firmly
in the LBL region, unlike ours, so that they should not have been expected to
show high $\nu_{\rm peak}$ values and steep {\it ASCA} spectra.

As shown in \S~\ref{peak}, the $\nu_{\rm peak}$ distributions are continuous
and peak at $\approx 10^{14}$ Hz for both FSRQ and BL Lacs. While on this
basis there appears to be no need for dividing blazars into ``H''
(high-energy) and ``L'' (low-energy) subclasses, the picture that comes out
for FSRQ is the same as that already known for BL Lacs (e.g., \citet{pad96}),
that is one of a division which can be based on physical grounds with ``H''
sources as those whose X-ray band is dominated by synchrotron emission and
``L'' sources as those in which inverse Compton dominates.

\subsection{Synchrotron Peak Frequency vs. Effective Spectral Indices}
\label{arx_peak}

It has been suggested in the literature (e.g., Padovani \& Giommi 1995b, 1996;
Fossati et al. 1998) that the blazar synchrotron peak frequency can be
estimated from the values of the effective spectral indices. We address
this point here by using for the first time a large, homogeneous and well-defined 
sample of blazars.

Figure \ref{arx_nubreak} plots $\alpha_{\rm rx}$ vs. $\nu_{\rm peak}$. As can
be seen, the two parameters correlate quite well (the correlation is
significant at the $> 99.9\%$ level for the whole sample and the FSRQ and at
the 98\% level for BL Lacs). However, there is considerable scatter - for any
given value of $\alpha_{\rm rx}$ the scatter in $\nu_{\rm peak}$ is more than
a decade. This is much more than reported in this relationship by
\citet{fos98}, whose Figure 8 shows a correlation with typically less than a
decade of scatter, particularly for $\nu_{\rm peak} \ga 10^{14}$ Hz. A careful
comparison shows that the major difference between the two plots occurs at low
$\alpha_{\rm rx}$. The plot in Fossati et al. shows that for $\nu_{\rm peak}
\la 10^{15.3}$ Hz there are no objects with $\alpha_{\rm rx} \la 0.75$, while
for $\nu_{\rm peak} \ga 10^{15.3}$ Hz the exact opposite is the case. We have
drawn these regions on Figure \ref{arx_nubreak} (dotted horizontal and
vertical lines). With DXRBS, however, we see a different picture. As Figure
\ref{arx_nubreak} shows, DXRBS has effectively expanded the accessible region
of parameter space by making accessible the region $0.65 \la \alpha_{\rm rx}
\la 0.75$ in the bottom left quadrant. The resulting increase in accessible
parameter space is huge: while for objects with $10^{12} \la \nu_{\rm peak}
\la 10^{13}$ Hz this is only about 20\%, it is more than 50\% for objects with
$10^{14} \la \nu_{\rm peak} \la 10^{15}$ Hz. A look at Figure
\ref{arx_nubreak} shows that while the correlation is still highly
significant, the plot looks very different at the high $\alpha_{\rm rx}$ end
than it does at low $\alpha_{\rm rx}$. At high $\alpha_{\rm rx}$ we see a
clear envelope, while it is obvious that we do not see an envelope at low
$\alpha_{\rm rx}$. In other words a DXRBS blazar with $\alpha_{\rm rx} \sim
0.7$ is just as likely to have $\nu_{\rm peak} \sim 10^{12}$ Hz as it is to
have $\nu_{\rm peak} \sim 10^{16}$ Hz. We note that the sources with
relatively low $\alpha_{\rm rx}$ {\it and} $\nu_{\rm peak}$ values are the
DXRBS sources in the upper-left part of Fig. \ref{aroaox} which have low
$\alpha_{\rm rx}$ but are outside of the HBL box.

What is the reason behind this difference? In the first place the models are
different: \citet{fos98} used a simple third-degree polynomial fit, while the
model we use is based upon a calculation of synchrotron and inverse-Compton
spectra for reasonable physical parameters. More importantly, however, all 
the samples used by Fossati et al. - the 1 Jy, 2 Jy and Slew surveys - are
classic, high flux limit surveys in the radio and X-rays respectively. As
shown in \S~\ref{discovery}, this type of survey has the effect of imposing
what amounts to severe selection biases. This can in fact be seen in Figure 8
of \citet{fos98}, where both their $\nu_{\rm peak} - \alpha_{\rm rx}$ and
$\nu_{\rm peak} - \alpha_{\rm ro}$ plots are seen to be discontinuous, hardly
surprising considering that all but one of the objects with $\nu_{\rm peak} >
10^{15.3}$ Hz in the Fossati et al. study is found in the Einstein Slew
Survey but not the 1 Jy (only 2/34 1 Jy BL Lacs fall in the lower right hand
quadrant of our Figure \ref{arx_nubreak} as denoted by the two dotted lines,
compared to 50/60 Slew BL Lacs), while the vast majority of the objects with
$\nu_{\rm peak}< 10^{15.3}$ Hz come from the 1 Jansky (33/34 1 Jy BL Lacs are
in this range compared to 10/60 Slew BL Lacs). 

\centerline{\includegraphics[width=9.0cm]{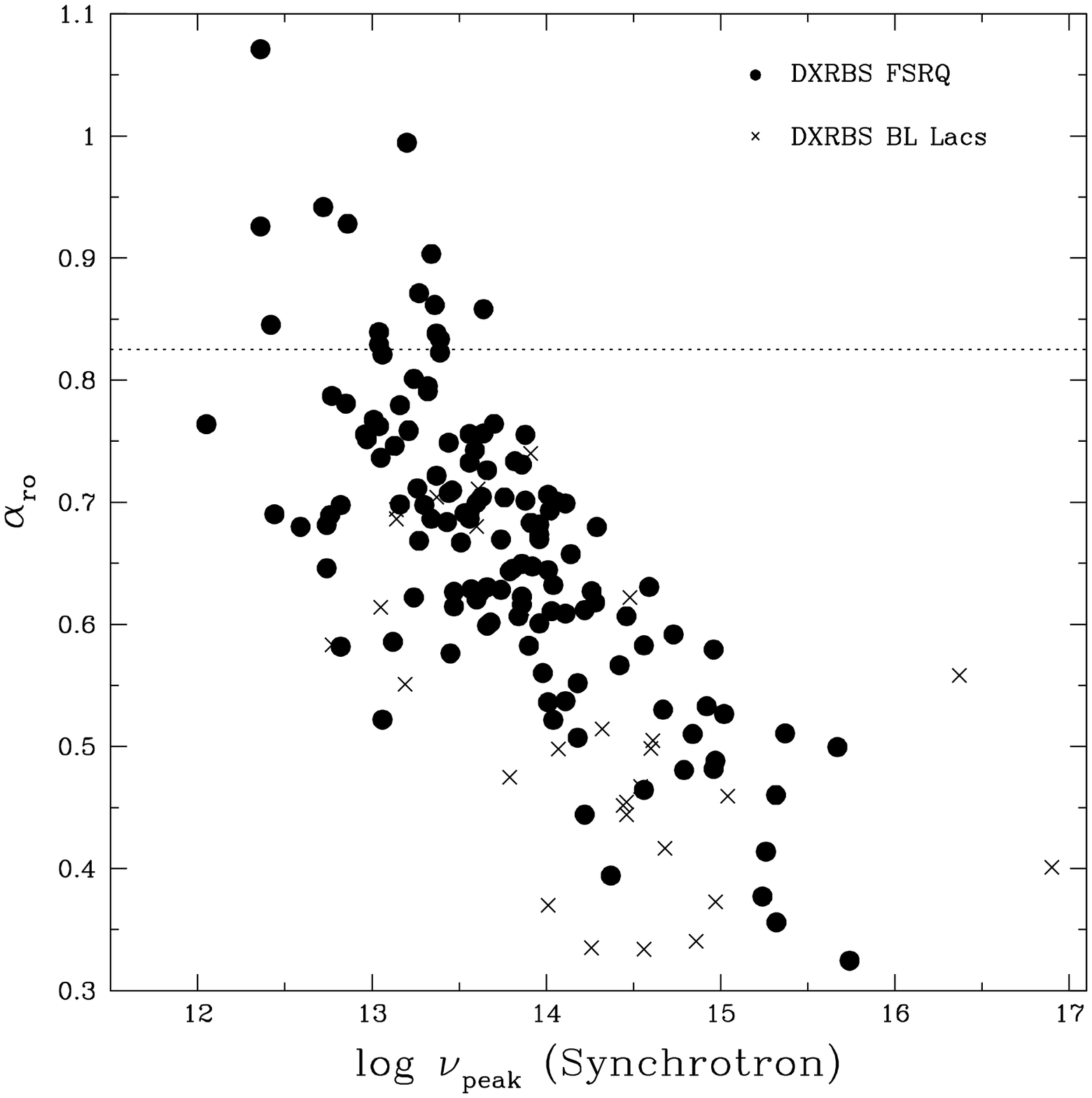}}
\figcaption{Radio-optical spectral index $\alpha_{\rm ro}$ vs. synchrotron peak
frequency $\nu_{\rm peak}$ for FSRQ (filled points) and BL Lacs (crosses) for
the DXRBS sample. The dotted line at $\alpha_{\rm ro} = 0.825$ denotes the
limit of the sources studied by \citet{fos98}. See text for
details.\label{aro_nubreak}}
\centerline{}
\vskip .2in

A tighter correlation appears to be present between $\alpha_{\rm ro}$ and
$\nu_{\rm peak}$, shown in Fig. \ref{aro_nubreak}. The correlation is
significant at the $> 99.99\%$ level ($\sim 97\%$ for BL Lacs only), with
$\alpha_{\rm ro} \propto \nu_{\rm peak}^{-0.12}$. This correlation is
expected to break down for $\nu_{\rm peak} \ga 10^{16}$ Hz or $\alpha_{\rm ro}
\la 0.4$ \citep{pad95b,fos98} but most of our sources are outside of this
range, which explains why the correlation looks almost linear. Apart from the
lack of objects with $\alpha_{\rm ro} \ga 0.8$ and the paucity of objects with
intermediate values of $\nu_{\rm peak}$ Fig. 8 of \citet{fos98} does not look
too different from Fig. \ref{aro_nubreak} in the overlapping range of
$\nu_{\rm peak}$. 

We notice that Fig. \ref{arx_nubreak} shows that a definition based solely on
X-ray-to-radio flux ratio or $\alpha_{\rm rx}$ is not optimal, as we have
found blazars with low $\alpha_{\rm rx}$ {\it and} low $\nu_{\rm peak}$
values. The position of the sources on the $\alpha_{\rm ox}$, $\alpha_{\rm
ro}$ plane, which means using two (instead of one) effective spectral indices,
on the other hand, appears to be more sensitive to the synchrotron peak
frequency, with a difference of a factor $\sim 35$ in the mean $\nu_{\rm
peak}$ values for blazars in and out of the HBL box. This is true especially
for FSRQ. Fig. \ref{aroaox} shows, in fact, that most of the blazars with
$\alpha_{\rm rx} \le 0.78$ but outside the HBL box are strong-lined. Tab. 1
shows also that, taking into account the sky coverage, while $\sim 54\%$ of BL
Lacs with $\alpha_{\rm rx} \le 0.78$ are in the HBL box, this is true only for
$\sim 36\%$ of the FSRQ.

\subsection{Other Properties of HBL/LBL and HFSRQ/LFSRQ}

We have studied the properties of blazars in and out of the HBL box not addressed
in \S~\ref{SED}, namely radio, optical, and X-ray powers and redshift. As
before, distributions are ``corrected'' by using the sky coverage. HFSRQ have
slightly smaller radio powers and slightly larger X-ray powers than LFSRQ, by
factors $\sim 2$, but not significantly so ($P \sim 90\%$). HFSRQ, on the
other hand, have significantly ($P > 99.99\%$) larger (factor $\sim 7$)
optical powers, due to the fact that for these sources $\langle \nu_{\rm peak}
\rangle \sim 7 \times 10^{14}$ Hz. Both classes have similar mean redshifts,
$\sim 1.6$ and $\sim 1.7$ respectively, with redshift distributions which are
different at the $96\%$ level. The same comparison in the case of BL Lacs is
hampered by the relatively small number statistics (18 LBL and 4 HBL with
redshift). With this caveat, we find that HBL have slightly larger radio and
optical powers than LBL, by factors $\sim 2 - 5$, but not significantly so ($P
\sim 50\%$ and $\sim 90\%$ respectively), while they have significantly ($P
\sim 98.7\%$) larger (factor $\sim 35$) X-ray powers. The redshift
distributions for the two classes are similar with $\langle z \rangle \sim
0.4$.

If indeed all blazars that lie within the HBL ``box'' have radio-to-X-ray
continua produced by synchrotron emission, one might expect a significant
population of HFSRQ to be 100 GeV -- TeV emitters. An example of such an
object might be RGB J1629+4008 \citep{pad02}, which has a synchrotron peak
$\sim 2 \times 10^{16}~{\rm Hz}$ and a predicted $\nu F_\nu$ at $10^{25}$ Hz
of nearly $10^{-10} {\rm ~erg ~cm^{-2} ~s^{-1}}$. A similar point was also
made by \citet{per00}, albeit without the aid of BeppoSAX spectral data, who
also listed five FSRQ with $\alpha_{\rm rx} <0.78$ and $z\leq0.1$ derived from
the RGB and {\it Einstein} Slew Survey. Thus the discovery of a new, X-ray
loud population of FSRQ predicts the existence of a new class of 100 GeV --
TeV emitting sources which could be particularly helpful for probing the
diffuse infrared background at higher redshifts.

\section{Summary and Conclusions}

We have used the results of two recent surveys, DXRBS and RGB, to study the
spectral energy distribution of about 500 blazars. Never before had this been
done with a sample even remotely close to ours in terms of size, depth, and
well-defined selection criteria. DXRBS, in particular, is $\sim 95\%$ complete
and reaches fluxes $\sim 20$ times lower than previously available blazar radio
surveys. We have first derived the effective spectral indices $\alpha_{\rm
ox}$, $\alpha_{\rm ro}$, and $\alpha_{\rm rx}$. We have then studied the
$\alpha_{\rm rx}$ distributions for DXRBS, RGB, and even the EMSS samples for
both FSRQ and BL Lacs. The serendipitous nature of DXRBS and the EMSS has been
taken into account by ``correcting'' these and other distributions using the
appropriate sky coverage. The synchrotron peak frequencies for DXRBS blazars
have also been derived by using multi-frequency information and an homogeneous
synchrotron - inverse self-Compton model. Broad line region and jet powers
were also estimated. One of the main aims of this work was to look for the
strong-lined counterparts of high-energy peaked (HBL) BL Lacs, the HFSRQ, that
is FSRQ with high-energy synchrotron peaks. We have found them. Our main
results can be summarized as follows:

\begin{enumerate} 
\item About $10\%$ of DXRBS FSRQ have effective spectral indices typical of
HBL (to be compared with $15\%$ for BL Lacs) and can therefore be called
HFSRQ. The fractions of HFSRQ and HBL increase to $\sim 30\%$ and $\sim 60\%$
for RGB and to $\sim 80\%$ and $100\%$ for the EMSS, respectively. Although
HFSRQ have X-ray-to-radio flux ratios larger than previously known FSRQ, in
none of the samples they manage to reach values as high as those of HBL.

\item The synchrotron peak frequency distribution of DXRBS FSRQ and BL Lacs is
continuous and peaks at $\approx 10^{14}$ Hz, with the former sources having
an average $\nu_{\rm peak} \sim 2.5$ times smaller than the latter. We have 
verified
that blazars with effective spectral indices typical of HBL indeed have larger
$\nu_{\rm peak}$ values (by a factor $\sim 35$) than other blazars. About
$60\%$ and $>8\%$ of DXRBS BL Lacs have $\nu_{\rm peak} > 10^{14}$ and
$10^{15}$ Hz respectively, to be compared with $\sim 30\%$ and $\sim 5\%$ for
FSRQ.

\item These results, together with the dependence we find of the X-ray
spectral index, estimated from the hardness ratios, on $\nu_{\rm peak}$, {\it
confirm the existence of strong-lined counterparts of high-energy peaked BL
Lacs}. As is the case for HBL, we would expect a significant fraction of these
sources to emit at 100 GeV -- TeV energies.

\item We find no anti-correlation between synchrotron peak frequency and
radio, broad line region, and jet powers, contrary to the predictions of the
so-called ``blazar sequence'' scenario, which calls for an inverse dependence
of $\nu_{\rm peak}$ on intrinsic power due to the effects of the more severe
electron cooling in more powerful sources. On the other hand, available data
from DXRBS and other surveys suggest that high-$\nu_{\rm peak}$--high-power
blazars have not been found yet, and that HFSRQ do not reach the extreme
synchrotron peak frequencies of BL Lacs. This indicates that after all there
might be an intrinsic, physical limit to the synchrotron peak frequencies and
therefore electron energies which can be reached by powerful blazars.
\end{enumerate}  

The discovery of HFSRQ and the study of their properties have important
implications for our understanding of jet formation and physics. Since
$\nu_{\rm peak} \propto \gamma^2_{\rm peak} \delta B$, where $\gamma_{\rm
peak}$ is the Lorentz factor of the electrons emitting most of the radiation,
we have shown that {\it powerful} jets with large magnetic fields {\it and}
electron Lorentz factors can indeed exist -- regardless of whether or not they
have strong emission lines -- albeit up to a point. This provides an important
challenge for existing models that advocate that the spectral energy
distribution of relativistic jets is strongly affected by the external
radiation field. We have also shown that selection effects are very strong and
that, in particular, the HBL/HFSRQ fraction is sample-dependent. Can we
separate selection effects from physics? We think we can. By using DXRBS we
have sampled a region of parameter space which should be largely unbiased in
terms of $\nu_{\rm peak}$ (unlike that covered by the EMSS, for example). We
have found that HBL/HFSRQ make up a minority of the blazar population, $\sim
10-15\%$. We then believe that the available evidence suggests that nature
preferentially makes jets which peak at IR/optical energies. We will address
this issue in detail in a future paper.

It is also clear that, although a consistent picture comes out of our results,
we need more information on these sources. We are planning dedicated X-ray
observations of our newly discovered HFSRQ to further constrain their X-ray
emission processes.

\acknowledgments

EP acknowledges support from NASA grants NAG5-9995 and NAG5-10109 (ADP) and
NAG5-9997 (LTSA). HL acknowledges financial support from the Deutscher
Akademischer Austauschdienst (DAAD) and the STScI DDRF grants D0001.82260 and
D0001.82299. This research has made use of the NASA/IPAC Extragalactic
Database (NED), which is operated by the Jet Propulsion Laboratory, California
Institute of Technology, under contract with the National Aeronautics and
Space Administration, and of data products from the Two Micron All Sky Survey,
which is a joint project of the University of Massachusetts and the Infrared
Processing and Analysis Center/California Institute of Technology, funded by
the National Aeronautics and Space Administration and the National Science
Foundation.

\end{document}